\documentclass[unnumsec,webpdf,modern,large]{oup-authoring-template}

\setcitestyle{round,authoryear}
\usepackage{graphicx}
\usepackage{booktabs}
\usepackage{amsmath,amssymb}
\usepackage[table]{xcolor}
\usepackage[most]{tcolorbox}   
\usepackage{longtable}
\usepackage{pifont}
\usepackage{balance}
\definecolor{cWin}{HTML}{1B6E2E}
\definecolor{cTie}{HTML}{555555}
\definecolor{cLoss}{HTML}{B83227}
\definecolor{cAbC}{HTML}{1B4F72}
\definecolor{cAbT}{HTML}{B7791F}
\definecolor{cAbA}{HTML}{6A2C70}
\definecolor{cAbD}{HTML}{0E6F6F}
\definecolor{cCardFrame}{HTML}{B3B3B3}
\definecolor{cCardTitleBg}{HTML}{F4F4F4}
\definecolor{cCodeBg}{HTML}{FAFAFA}
\definecolor{cBlockBg}{HTML}{F4F8F8}
\definecolor{cMuted}{HTML}{777777}
\definecolor{cRouteBg}{HTML}{FFF4F7}

\providecommand{\ablBadge}[2]{%
  {\setlength{\fboxsep}{1.7pt}%
   \colorbox{#1}{\color{white}\scriptsize\bfseries\strut\,#2\,}}}
\providecommand{\verdictBadge}[2]{%
  {\setlength{\fboxsep}{1.5pt}%
   \fcolorbox{#1}{white}{\color{#1}\scriptsize\bfseries\strut\,#2\,}}}

\graphicspath{{Fig/}}

\theoremstyle{thmstyleone}%
\theoremstyle{thmstyletwo}%
\theoremstyle{thmstylethree}%

\newcommand{\model}{{BioHarness}}

\begin{document}

\journaltitle{Bioinformatics}
\DOI{DOI HERE}
\copyrightyear{2022}
\pubyear{2019}
\access{Advance Access Publication Date: Day Month Year}
\appnotes{Paper}

\firstpage{1}


\title[\model]{\model: Substrate-Aware Evidence Assembly for Biomedical Question Answering across Literature, Knowledge Bases, and Biological Atlases}

\author[1,2]{Meng Xiao\ORCID{0000-0001-5294-5776}}
\author[1,3]{Chuan Qin\ORCID{0000-0002-5354-8630}}
\author[2,4]{Jinmiao Chen\ORCID{0000-0001-7547-6423}}
\author[1]{Yihang Cheng\ORCID{0000-0003-2396-0306}}
\author[1,3,$\ast$]{Yuanchun Zhou\ORCID{0000-0002-1987-2402}}
\author[1,3,$\ast$]{Hengshu Zhu\ORCID{0000-0003-4570-643X}}

\authormark{Meng Xiao et al.}

\address[1]{\orgdiv{Computer Network Information Center}, \orgname{Chinese Academy of Sciences}, 
\orgaddress{\street{South Fourth Street}, \postcode{100180}, \state{Beijing}, \country{China}}}
\address[2]{\orgdiv{DUKE-NUS Medical School}, \orgname{National University of Singapore}, \orgaddress{\street{8 College Rd}, \postcode{169857}, \country{Singapore}}}
\address[3]{\orgname{University of Chinese Academy of Sciences}, \orgaddress{\street{Yanqihu East Rd}, \postcode{101408}, \country{China}}}
\address[4]{\orgdiv{Bioinformatics Institute}, \orgname{Agency for Science, Technology and Research}, \orgaddress{\street{30 Biopolis Street}, \postcode{138671}, \country{Singapore}}}


\corresp[$\ast$]{Corresponding authors: Yuanchun Zhou (\href{zyc@cnic.cn}{zyc@cnic.cn}) and Hengshu Zhu (\href{hszhu@cnic.cn}{hszhu@cnic.cn})}

\received{Date}{0}{Year}
\revised{Date}{0}{Year}
\accepted{Date}{0}{Year}



\abstract{
\textbf{Motivation}:
Biomedical question answering often requires evidence beyond topically retrieved literature, including gene alias resolution, database identifier normalization, and atlas-derived biological measurements.
However, existing retrieval-augmented generation (RAG) systems typically follow a fixed workflow and lack an explicit mechanism for deciding when retrieved text is sufficient, when curated biomedical knowledge is required, or when executable evidence assembly over structured measurements should be invoked.
This motivates a substrate-aware large language model (LLM) harness that selectively assembles sufficient evidence across literature, knowledge bases, and biological atlases.
\\
\textbf{Results}:
We introduce \model{}, an LLM harness for staged biomedical evidence assembly across literature retrieval, curated biomedical knowledge resources, and atlas-derived structured measurements.
\model{} first attempts to answer from reranked literature evidence, and escalates through grounded cascade control to REPL-style evidence assembly only when the current evidence is uncertain, weakly grounded, or substrate-mismatched.
Across 19,302 biomedical QA items spanning seven answer formats, \model{} improves the pooled score from 65.9 to 71.0 over the strongest non-oracle baseline.
Ablations, case studies, and backbone-scaling analyses show that these gains arise from repairing evidence-substrate mismatches through reranking, entity grounding, and structured measurement access, rather than from indiscriminately invoking more reasoning steps, retrieving additional literature, or relying on a particular answer-model scale. 
\\
\textbf{Availability and implementation}:
Source code\footnote{\url{https://github.com/coco11563/BioHarness_Chi}}, reproducible benchmarking scripts\footnote{\url{https://github.com/coco11563/BioHarness_Eval_Framework}}, and processed data\footnote{\url{https://huggingface.co/datasets/Shaow/BioHarness_Eval}} are all publicly available.
} 
\keywords{biomedical question answering; retrieval-augmented generation; large language models; evidence assembly; biomedical knowledge bases; biological atlases}

\maketitle

\section{Introduction}
\begin{figure}
\centering
\includegraphics[width=\linewidth]{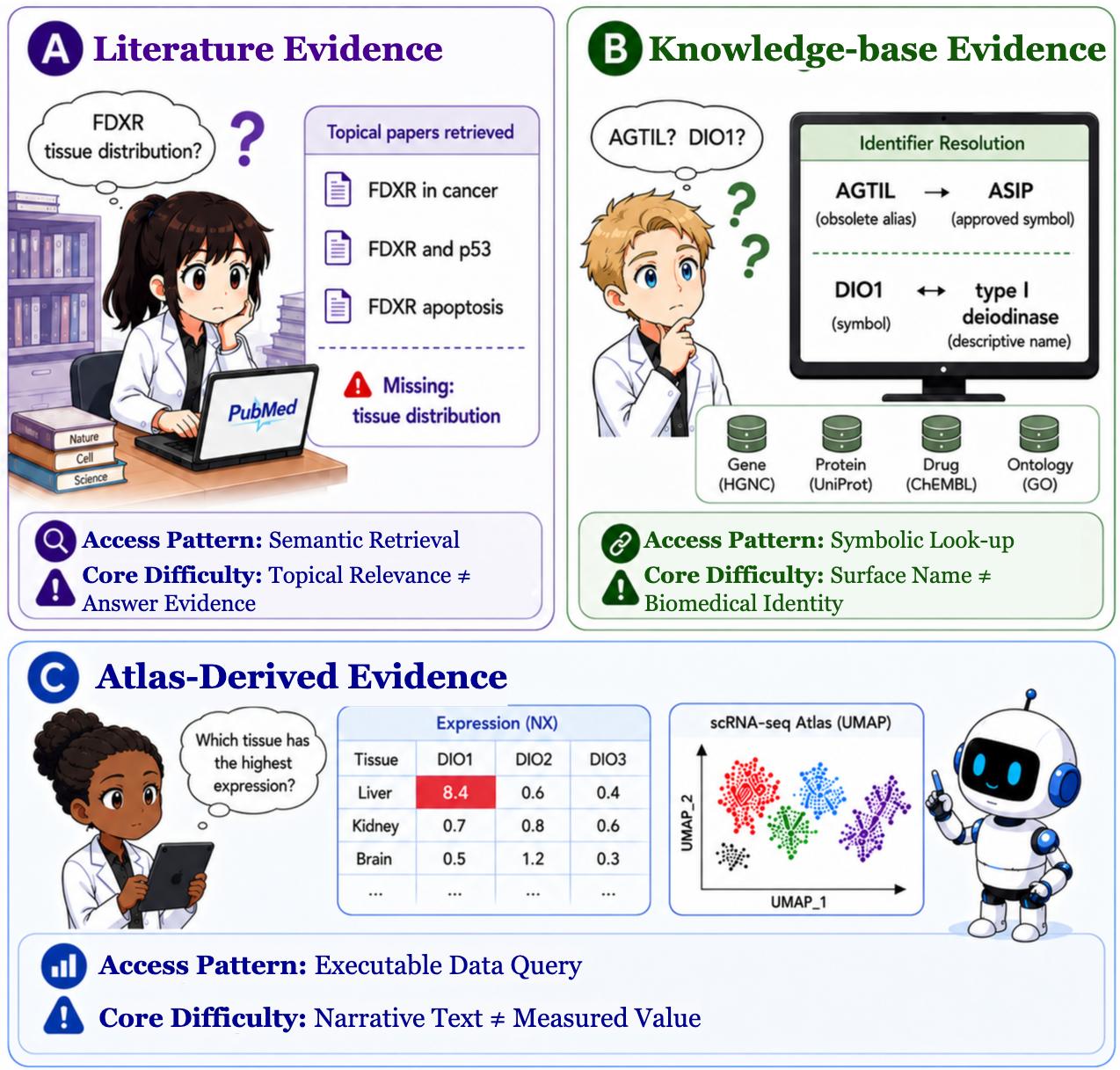}
\caption{Heterogeneous evidence substrates in biomedical question answering. Literature evidence supports claim-level reasoning through semantic retrieval, knowledge-base evidence supports entity-level grounding through identifier and relation lookup, and atlas-derived evidence supports measurement-level reasoning through executable queries over structured biological data.
}
\label{fig:intro}
\end{figure}

Biomedical question answering with large language models~\citep{luo2025large,zhou2025large,zhang2026data} is not a single-substrate retrieval problem. 
A biomedical question may require an answer-bearing literature statement~\citep{jin2022biomedical,jin2023medcpt}, canonical entity grounding behind an ambiguous surface mention~\citep{wei2024pubtator,gyori2022gilda}, or a measured expression pattern stored in a bulk expression resource, tissue atlas, or single-cell atlas~\citep{george2024expression,li2022disco,the2022tabula,czi2025cz}.
We refer to these distinct forms of evidence as biomedical evidence substrates because they differ not only in content but also in access patterns, reliability profiles, and failure modes.
Literature evidence is accessed via semantic text retrieval, knowledge-base evidence via identifier and relation lookup, and atlas-derived evidence via executable operations on structured biological measurements.
As illustrated in Figure~\ref{fig:intro}, fixed retrieve-then-generate pipelines can fail through evidence-substrate mismatch: topically relevant text may not contain answer-supporting evidence, surface names may not resolve to the correct biomedical identity, and measured biological values may be absent from narrative abstracts even when they are available in structured resources.

Existing biomedical RAG systems improve individual parts of this process, but rarely treat evidence-substrate selection as an explicit control problem. 
Text-centered methods, including dense retrieval, query expansion, reranking, and reflective evidence filtering, improve the recall or ranking of relevant passages, but still assume that the answer-supporting evidence can be recovered from retrieved text~\citep{lewis2020retrieval,karpukhin2020dense,gao2023precise,yu2024rankrag,asai2023self,yu2024chain}.
Biomedical tool-augmented and knowledge-graph-based methods incorporate structured resources or entity-relation context, yet these resources are often attached as fixed preprocessing, auxiliary context, or graph summaries rather than invoked according to evidence sufficiency~\citep{jin2024genegpt,wang2024biorag,li2025biomedrag,edge2024local,matsumoto2024kragen,soman2024biomedical}.
Agentic retrieval and reasoning frameworks can further interleave search, tool use, and multi-step inference, but indiscriminate execution increases costs and may still operate on the wrong substrate when the question requires canonical entity grounding or database-native measurements~\citep{yao2022react,trivedi2023ircot,shinn2023reflexion}.
Thus, the central problem is not simply how to retrieve more passages, add more tools, or reason for more steps, but how to decide which evidence substrate is sufficient for a given biomedical question.

To address this limitation, we introduce \model{}, a substrate-aware LLM harness for staged biomedical evidence assembly and question answering.
\model{} first constructs a literature evidence context through complementary retrieval views and question-conditioned reranking, then applies grounded cascade control to decide whether the current evidence is sufficient for a fast answer.
When the retrieved evidence is uncertain, weakly grounded, or mismatched to the question requirement, \model{} escalates to a REPL-style evidence assembly loop, where the LLM can inspect passages, invoke biomedical tools for entity grounding, access router-gated atlas evidence, and assemble a compact evidence state for final answer generation.
A task-constrained output head finally normalizes the answer to the required format, allowing the same harness to support yes/no, multiple-choice, factoid, list, summary, and expression-style questions.

Our contributions are threefold.
First, we identify evidence-substrate mismatch as a central failure mode in biomedical RAG, showing that failures can arise not only from insufficient retrieval but also from applying textual retrieval to questions that require entity grounding or structured biological measurements.
Second, we propose \model{}, a substrate-aware LLM harness that connects literature evidence, curated biomedical knowledge, and atlas-derived biological measurements through dual evidence retrieval, question-conditioned reranking, cascade control, biomedical tools, router-gated atlas access, REPL-style evidence repair, and task-constrained answer generation.
Third, we conduct an extensive controlled analysis across heterogeneous biomedical QA datasets, component ablations, efficiency settings, backbone scales, and mechanistic case studies, demonstrating that substrate-aware evidence assembly improves biomedical QA beyond fixed retrieval, query-expansion, reflective, tool-augmented, and graph-based RAG pipelines.

\section{Method}
\begin{figure*}
    \centering
    \includegraphics[width=\linewidth]{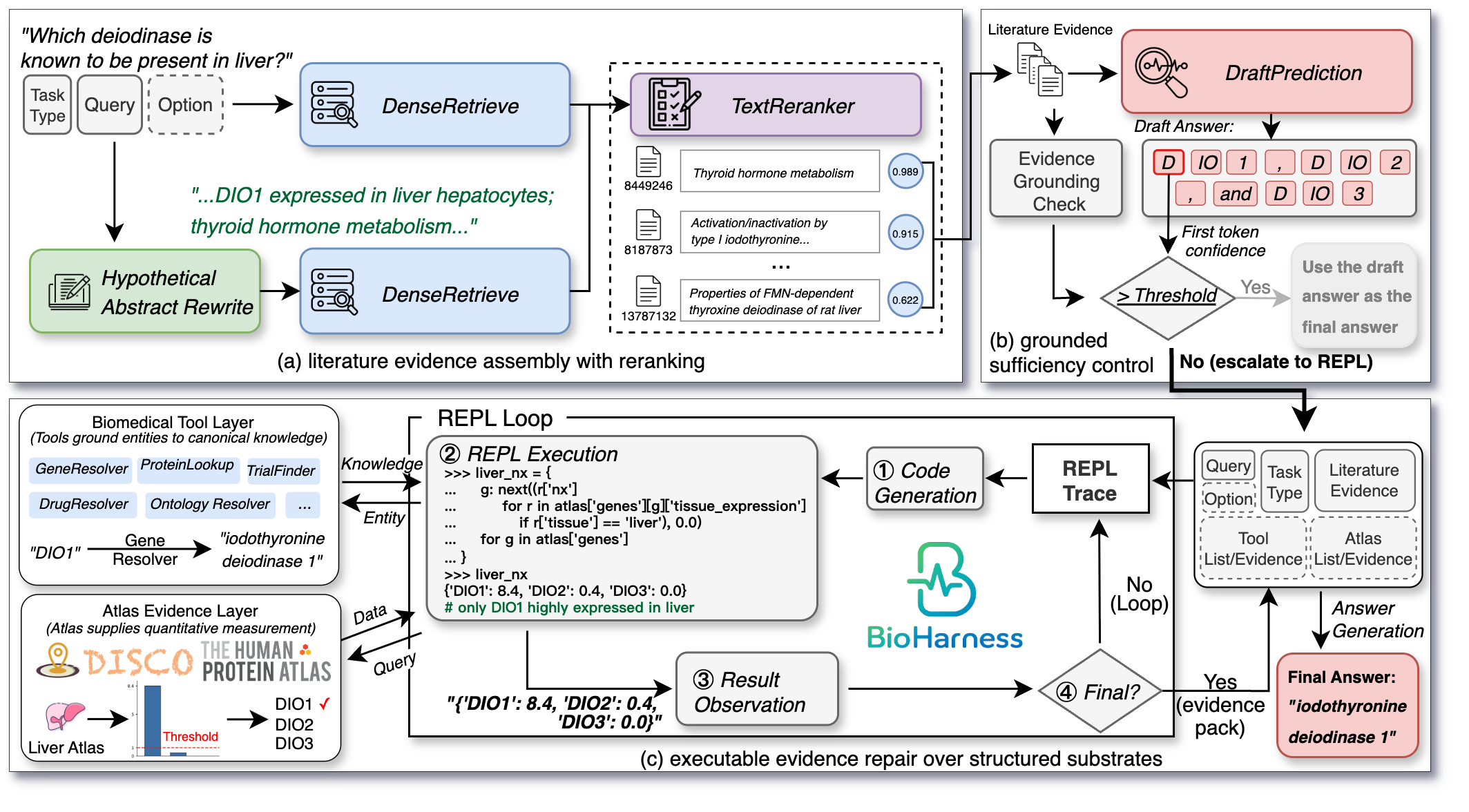}
    \vspace{-1cm}
    \caption{\model{} implements a substrate-aware biomedical evidence assembly pipeline consisting of (a) literature evidence assembly with reranking, (b) grounded evidence sufficiency control, and (c) executable evidence repair over biomedical tools and atlas-derived structured measurements, followed by a task-constrained answer generation head.}
    \label{fig:overview}
\end{figure*}

As shown in Figure~\ref{fig:overview} and Algorithm~\ref{alg:cascade}, \model{} implements a substrate-aware biomedical evidence assembly pipeline that first attempts to answer from literature evidence and escalates only when additional evidence substrates are needed.  
When the literature context is uncertain, weakly grounded, or substrate-mismatched, REPL-style evidence repair incorporates biomedical tool outputs and atlas-derived measurement records to construct a sufficient evidence state.

\subsection{Literature Evidence Assembly}\label{sec:dual-retrieval}
The literature evidence component constructs an initial evidence context from a biomedical literature corpus constructed from PubMed abstracts.
This component addresses a basic limitation of single-query dense retrieval: top-ranked abstracts may be topically related to the question, but still miss the specific answer-bearing evidence needed by the downstream task.
Within the literature substrate, \model{} combines complementary retrieval views with question-conditioned reranking to improve recall while preserving answer relevance.

\begin{figure}[t]
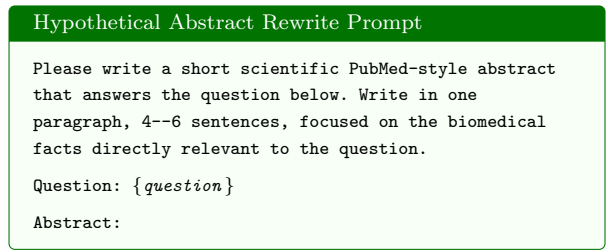

\centering
\begin{tcolorbox}[
title={Hypothetical Abstract Rewrite Prompt},
width=0.92\linewidth,
colback=green!3,
colframe=green!45!black,
boxrule=0.4pt,
arc=2pt,
left=6pt,
right=6pt,
top=5pt,
bottom=5pt
]
\small
\ttfamily
Please write a short scientific PubMed-style abstract that answers the
question below. Write in one paragraph, 4--6 sentences, focused on the
biomedical facts directly relevant to the question.

\vspace{4pt}

Question: \{\textit{\textbf{question}}\}

\vspace{4pt}

Abstract:
\end{tcolorbox}

\caption{
Prompt used for Hypothetical Abstract Rewrite. The biomedical question is converted into a hypothetical abstract and then used as an additional retrieval query.
}
\vspace{-1cm}
\label{prompt}
\end{figure}

As shown in Figure~\ref{fig:overview}(a), the first retrieval view uses the original question, preserving literal wording and entity-level precision.
The second retrieval view follows the hypothetical-document retrieval principle~\citep{gao2023precise}: the LLM rewrites the question as a short PubMed-style hypothetical abstract, as in the prompt in Figure~\ref{prompt}, thereby moving the retrieval query closer to a plausible biomedical evidence context.
The retrieved candidates from active retrieval views are merged, de-duplicated by document identifier when available, and reranked against the original question.
This reranking step mitigates hypothesis drift from the rewritten query while retaining its recall benefit.

The resulting Literature Evidence context is passed to the cascade controller as the initial evidence substrate.
Importantly, this component is not intended to solve all biomedical evidence needs through literature retrieval alone.
When a question requires contrastive evidence, the literature retriever may include task-aware query variants before reranking; when the primary evidence need is canonical entity grounding or structured biological measurement, the retrieved literature is retained as a lightweight context and the unresolved evidence need is delegated to downstream structured evidence substrates.
Thus, literature evidence assembly serves as the first substrate in a staged evidence assembly process, rather than as a universal retrieval solution.

\subsection{Grounded Evidence Sufficiency Control}

After constructing the Literature Evidence context, \model{} applies a grounded evidence sufficiency controller to decide whether the current evidence is adequate or whether additional evidence substrates should be invoked.
The controller is designed to separate answer generation from evidence sufficiency estimation: literature evidence is used when it provides a confident, well-grounded answer, whereas structured evidence repair is invoked when the draft answer is uncertain, unsupported, or mismatched with the question's requirements.
As shown in Figure~\ref{fig:overview}(b) and Algorithm~\ref{alg:cascade}, given the literature evidence set $\mathcal{D}$, \model{} first generates a draft answer $a^{(0)}$ under the task format and computes a lightweight confidence score $c=\exp(\ell)$, where $\ell$ is the log-probability of the first generated token.
This confidence score is used only as a routing proxy, rather than as a calibrated estimate of answer correctness~\citep{kadavath2022language}.
In parallel, the controller applies an answer-form-aware grounding check to estimate whether the draft is supported by the retrieved context.
For answer forms whose support can be localized in text, the check compares the normalized draft answer or its content tokens against the retrieved titles and abstracts; when lexical grounding is unreliable for a task format, routing relies primarily on the confidence signal and downstream evidence repair.

The controller escalates a query when the draft answer is empty, has low confidence, or is not grounded in the retrieved evidence.
If escalation is not triggered, the draft answer is returned as the final answer. 
Otherwise, the query is passed to REPL-style evidence repair, where biomedical tools and atlas-derived measurement records are initialized as structured evidence substrates.
This policy preserves the literature-only path when textual evidence is sufficient, while routing uncertain or weakly grounded drafts to the substrates required for evidence repair.
\begin{algorithm}[t]
\caption{Grounded Evidence Sufficiency Controller of \model.}
\label{alg:cascade}
\begin{algorithmic}[1]
\Require question $q$, task type $t$, options $O$ if applicable
\Ensure task-constrained answer $a$

\State $\mathcal{D} \gets \textsc{LiteratureEvidenceAssembly}(q,t)$
\State $(a^{(0)},\ell) \gets \textsc{DraftPrediction}(q,\mathcal{D};t,O)$
\State $c \gets \exp(\ell)$ \hfill \textcolor{cWin}{\# Confidence proxy}
\State $e \gets \textsc{EvidenceGroundingCheck}(a^{(0)},\mathcal{D},t)$
\hfill \textcolor{cWin}{\# Evidence grounding signal}

\State $\pi \gets \textsc{CascadeDecision}(t,c,e,a^{(0)};\theta_c)$
\If{$\pi=\textsc{Fast}$}
\State \Return $a^{(0)}$ \hfill \textcolor{cWin}{\# Literature-only path}
\EndIf

\State $h \gets \textsc{REPLEvidenceAssembly}(q,t,\mathcal{D},O;N_{\mathrm{iter}})$
\hfill \textcolor{cWin}{\# Algorithm~\ref{alg:repl}}
\State $a \gets \textsc{FinalAnswer}(q,\textsc{EvidencePack}(h,\mathcal{D});t,O)$
\State \Return $a$
\end{algorithmic}
\end{algorithm}

\subsection{Structured Biomedical Evidence Substrates}
\label{sec:tool_atlas}

When literature evidence is insufficient, \model{} can access two structured biomedical evidence substrates: the Biomedical Tool Layer and the Atlas Evidence Layer\footnote{The details of involving biomedical tools and atlas can be found in the attachment.}.
Together, they provide access patterns that are difficult to recover through dense retrieval alone.
The Biomedical Tool Layer supports entity-level grounding, including gene-symbol normalization, alias and deprecated-symbol resolution, identifier lookup, ontology traversal, and lightweight database-style checks.
These operations are treated as evidence-producing functions rather than external text generators: their role is to bridge ambiguous surface mentions to canonical biomedical entities, identifiers, and curated relations.

The Atlas Evidence Layer supports measurement-level grounding through a read-only atlas namespace.
Atlas access is controlled by a router that converts atlas-relevant questions into an atlas query specification, including candidate genes, tissues or cell types, measurement fields, and optional operations such as filtering or thresholding.
Only routed cases fetch the corresponding atlas entries before REPL execution.
The fetched atlas object is represented as structured measurement evidence, containing gene metadata, tissue-expression records, cell-type annotations, quantitative values, and provenance fields when available.
During REPL execution, these structured records can be inspected, filtered, aggregated, or thresholded through deterministic operations.

The two substrates are deliberately separated because they address different evidence needs.
Biomedical tools resolve names, identifiers, aliases, ontological terms, and curated database relations, whereas atlas access supplies biological measurements such as tissue-level or cell-type-level expression profiles.
This separation allows \model{} to switch evidence substrates according to the failure mode of the literature context, rather than repeatedly querying the same textual corpus.

\begin{algorithm}[t]
\caption{REPL-style Evidence Repair.}
\label{alg:repl}
\begin{algorithmic}[1]
\Require question $q$, task type $t$, literature evidence $\mathcal{D}$, options $O$ if applicable, iteration cap $N_{\mathrm{iter}}$
\Ensure evidence trace $h$

\Statex \textcolor{cMuted}{\# Stage A -- initialize heterogeneous evidence substrates}
\State $T \gets \textsc{InitBiomedicalTools}(q,t)$
\hfill \textcolor{cWin}{\# Initialize callable grounding tools}

\State $r \gets \textsc{AtlasRouter}(q,t)$
\hfill \textcolor{cWin}{\# Generate atlas query specification}

\If{$r.\mathrm{route_atlas}$}
\State $A \gets \textsc{FetchAtlasEvidence}(r.\mathrm{query_spec})$
\hfill \textcolor{cWin}{\# Structured measurement evidence}
\Else
\State $A \gets \varnothing$
\EndIf

\State $h \gets \textsc{InitContext}(q,t,\mathcal{D},T,A,O)$

\Statex \textcolor{cMuted}{\# Stage B -- iterative executable evidence repair}
\For{$i = 1,\ldots,N_{\mathrm{iter}}$}

\State $\mathrm{code}_i \gets \textsc{CodeGeneration}(h)$
\hfill \textcolor{cWin}{\# \ding{172} Code generation}

\State $\mathrm{exec}_i \gets \textsc{ReplExec}(\mathrm{code}_i)$
\hfill \textcolor{cWin}{\# \ding{173} REPL execution}

\State $\mathrm{obs}_i \gets \textsc{ResultObservation}(\mathrm{exec}_i)$
\hfill \textcolor{cWin}{\# \ding{174} Result observation}

\State $h \gets h\Vert \mathrm{code}_i \Vert \mathrm{obs}_i$
\hfill \textcolor{cWin}{\# Update evidence trace}

\If{$\textsc{IsFinal}(\mathrm{obs}_i)$}
    \State \textbf{break}
    \hfill \textcolor{cWin}{\# \ding{175} Stop if sufficient}
\EndIf

\EndFor

\State \Return $h$
\end{algorithmic}
\end{algorithm}

\subsection{Executable Evidence Repair}

When the controller rejects the literature-only path, \model{} enters a REPL-style evidence repair loop.
As shown in Algorithm~\ref{alg:repl}, the loop is not used as an unconstrained reasoning chain.
Instead, it provides an executable workspace for substrate-aware evidence repair: the LLM can inspect retrieved passages, invoke read-only biomedical tools, access router-gated atlas evidence, and assemble a compact evidence state through executable operations.

This design targets two common repair needs after literature retrieval.
First, some questions require symbolic biomedical grounding rather than additional text retrieval.
Deprecated aliases, surface-form ambiguity, and missing identifiers are better resolved by gene normalization and database lookup tools than by repeatedly searching the literature corpus.
Second, atlas-derived evidence is often structured and numeric rather than narrative.
Expression patterns, tissue distributions, and cell-type measurements naturally require executable operations such as filtering rows, applying thresholds, aggregating values, or constructing entity sets.
A REPL-style interface therefore matches the access pattern of structured atlas evidence more closely than a pure retrieve-then-generate pipeline.

The evidence trace $h$ records the action--observation history and is later compressed into an evidence pack for final answer generation.
The REPL interface gives the harness two properties that ordinary retrieve-then-generate pipelines lack.
First, observations are stateful: intermediate outputs such as resolved symbols, candidate entity sets, database records, tissue-expression rows, and filtered atlas measurements remain available for later operations.
Second, execution is inspectable: the trace exposes whether the repair came from entity normalization, database lookup, atlas filtering, or another substrate-specific operation.
This makes evidence repair both substrate-aware and auditable, rather than hidden inside free-form generation.

\subsection{Task-constrained Answer Generation}
\label{sec:answer-generation}

After either the literature-only path or REPL-style evidence repair, \model{} produces the final answer through a task-constrained output head.
For literature-only cases, the draft answer is returned when the controller determines that it is sufficiently confident and grounded in the literature evidence.
For escalated cases, the REPL trace is compressed into an evidence pack that includes retrieved literature snippets, resolved biomedical entities, tool observations, atlas-derived measurement records, and, when available, provenance information.
The output head then converts this evidence pack into an answer compatible with the task specification and answer options.

This step separates evidence assembly from answer formatting.
Instead of relying on unconstrained free-form generation, the output head constrains the response according to the required output space, such as closed-set options, normalized entity spans or entity sets, or concise evidence-grounded summaries.
This ensures that heterogeneous evidence traces from literature, biomedical tools, and atlas-derived measurements are converted into outputs compatible with the evaluation protocol.

\section{Experimental Settings}
\subsection{Task Description}

We evaluate biomedical QA in a unified evidence-assembly setting.
Given a query $q$, task type $t$, and answer options $O$ when available, the system is required to produce an answer $a$ using the assembled evidence context.
The benchmark covers heterogeneous answer formats, including yes/no questions, single-answer multiple-choice questions, multi-answer choice questions, short factoid answers, list-style answers, summaries, and structured expression outputs.

\subsection{Evaluation Metrics}
\label{sec:evaluation_metrics}

We evaluate each item using a type-specific matching score and report all scores on a 0--100 scale.
For yes/no questions, we use lowercased exact-match accuracy.
For single-answer multiple-choice questions, we use exact match on the predicted option letter.
For multi-answer choice questions, we compute set-F1 over the predicted and gold option sets.
For factoid questions, answers are usually short entities or phrases, so we use token-level F1~\citep{rajpurkar2016squad} after answer normalization, equivalent to unigram-overlap F1.
For list-style questions, we compute synonym-aware set-F1 between the predicted and gold answer sets.
For summary questions, we use ROUGE-L~\citep{lin2004rouge} F1 because it captures longest-common-subsequence overlap and better reflects content preservation in longer free-form answers.
For expression-style questions, we compute entity-set F1 over the target entity set.

Let $s_i \in [0,1]$ denote the type-specific matching score for item $i$.
For a dataset $\mathcal{D}$, the reported dataset score is computed as
\[
100 \times \frac{1}{|\mathcal{D}|}\sum_{i \in \mathcal{D}} s_i .
\]
The overall score is computed analogously over the full benchmark with size $N$:
\[
100 \times \frac{1}{N}\sum_{i=1}^{N} s_i,
\]
Thus, all reported scores are pooled item-level averages of type-specific matching scores.

\subsection{Implementation Details}

\begin{figure}[t]
\centering
\includegraphics[width=0.95\linewidth]{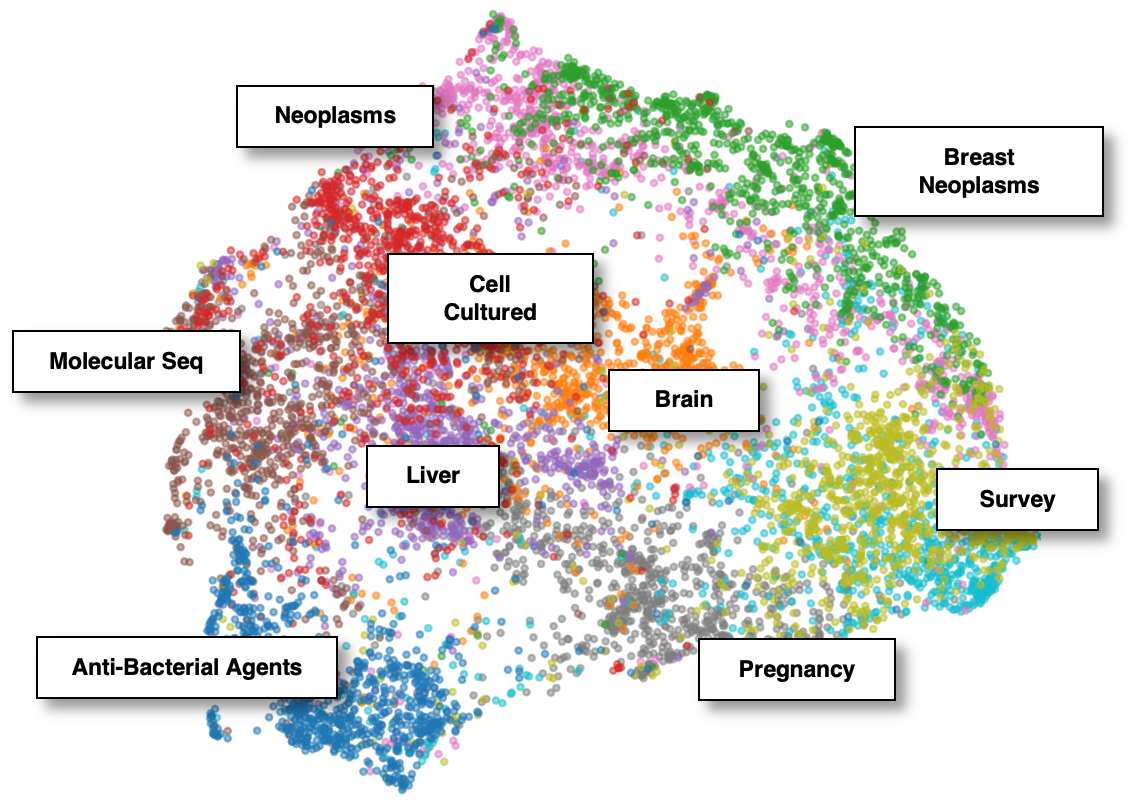}
\caption{MeSH-colored subsampled visualization of the PubMed embedding space. Each point denotes a PubMed abstract encoded by Qwen3-Embedding-0.6B, with colors assigned according to article-associated MeSH terms and labels marking representative MeSH categories.}
\label{fig:embedding_visualization}
\end{figure}

Experiments are conducted on a server equipped with eight NVIDIA A800-SXM4-80GB GPUs.
All LLMs are served locally using vLLM\footnote{\url{https://vllm.ai}}, with deterministic decoding and identical generation settings across methods.
Unless otherwise specified, generation-based components use Qwen3.5-35B-A3B\footnote{\url{https://huggingface.co/Qwen/Qwen3.5-35B-A3B}} as the common backbone, including hypothetical abstract rewriting, draft prediction, REPL-style evidence repair, and final answer generation. For literature retrieval, we construct a unified PubMed-derived corpus from approximately 20 million publicly available abstracts.
Questions and abstracts are encoded using Qwen3-Embedding-0.6B\footnote{\url{https://huggingface.co/Qwen/Qwen3-Embedding-0.6B}}, and abstracts are indexed in Qdrant\footnote{\url{https://qdrant.tech}} for dense vector retrieval.
Retrieved candidates are reranked under the original question using Qwen3-Reranker-8B\footnote{\url{https://huggingface.co/Qwen/Qwen3-Reranker-8B}}.
Figure~\ref{fig:embedding_visualization} provides a subsampled visualization of the PubMed abstract embedding space, in which MeSH-colored neighborhoods indicate that the retrieval corpus preserves the broad biomedical topic structure.
We use this visualization only as a qualitative sanity check of the literature retrieval substrate; all quantitative comparisons are reported in the benchmark results.
We set the grounding threshold $\theta_g$ to 0.5 and the confidence threshold $\theta_c$ to 0.7 in all experiments. 
Unless otherwise specified, all non-oracle baselines are implemented under the same retrieval corpus, embedding model, reranking model, answer backbone, decoding and formatting strategy, and evaluation pipeline.
This controlled setup ensures that performance differences primarily reflect evidence assembly and control strategies rather than platform-level differences.

\begin{table*}[t]
\centering
\setlength{\tabcolsep}{3.2pt}
\renewcommand{\arraystretch}{0.89}
\caption{
Main results on eight biomedical QA benchmarks.
\textit{Golden Context} uses dataset-provided gold evidence when available and is reported as an oracle evidence upper bound; it is excluded from best/second-best highlighting.
Among non-oracle methods, the best and second-best results are shown in \textbf{bold} and \underline{underlined}, respectively; ties are marked accordingly.
}
\label{tab:main_results}
\resizebox{\textwidth}{!}{
\begin{tabular}{lccccccccc}
\toprule
\textbf{Method}
& \textbf{PubMedQA}
& \textbf{BioASQ}
& \textbf{GeneTuring}
& \textbf{SciHorizon}
& \textbf{MedMCQA}
& \textbf{MedQA-US}
& \textbf{MedQA-TW}
& \textbf{MedQA-CN}
& \textbf{Overall} \\
\midrule

\rowcolor{gray!12}
\multicolumn{10}{c}{\textit{Retrieval-free baseline}} \\
\midrule
No-Context
& 50.6 & 45.4 & 7.9 & 42.1 & 72.7 & 82.0 & 89.1 & 89.3 & 62.1 \\
w/ Golden Context$^{\dagger}$
& 78.0 & 66.4 & -- & -- & -- & -- & -- & -- & -- \\

\midrule
\rowcolor{gray!12}
\multicolumn{10}{c}{\textit{Dense retrieval and query augmentation}} \\
\midrule
DenseRAG
& 75.4 & 47.2 & 8.9 & 43.3 & 73.1 & 84.0 & 89.0 & \underline{89.4} & 63.6 \\
HyDE
& 75.4 & 53.3 & 10.6 & 48.1 & 72.9 & \underline{84.1} & 88.5 & 88.8 & {65.7} \\

\midrule
\rowcolor{gray!12}
\multicolumn{10}{c}{\textit{Reranking, filtering, and reflective retrieval}} \\
\midrule
DenseRAG + Rerank
& 75.9 & 50.1 & 9.6 & 45.4 & \underline{73.5} & 82.6 & \underline{89.3} & 89.0 & 64.4 \\
Chain-of-Note
& 73.4 & 49.9 & 7.7 & 44.8 & 71.9 & 82.1 & 88.0 & 88.9 & 63.8 \\
Self-RAG & 74.0 & 48.7 & 9.1 & 45.1 & 73.1 & 81.9 & \textbf{89.5} & 89.2 & 64.1 \\
IRCoT
& \underline{76.0} & \underline{53.8} & 8.7 & 46.4 & 73.0 & 82.8 & 88.1 & 88.8 & 65.4 \\
\midrule
\rowcolor{gray!12}
\multicolumn{10}{c}{\textit{graph-based retrieval}} \\
\midrule
GraphRAG
& 50.8 & 44.9 & 8.7 & 44.7 & 72.0 & 82.2 & 87.8 & 89.0 & 62.1 \\
PathRAG
& 49.8 & 43.3 & 7.3 & 44.6 & 66.0 & 81.9 & 85.6 & 83.4 & 59.1 \\
LightRAG
& 57.0 & 44.6 & 9.0 & 44.2 & 71.6 & 82.2 & 88.3 & 88.8 & 62.1 \\
\midrule
\rowcolor{gray!12}
\multicolumn{10}{c}{\textit{Biomedical, tool-augmented retrieval}} \\
\midrule
FLARE & 72.4 & 39.2 & 9.7 & \underline{51.7} & 63.4 & 78.1 & 74.7 & 77.9 & 57.2 \\
BiomedRAG
& 75.0 & 47.8 & 9.5 & 44.8 & 72.9 & 82.4 & 88.3 & 89.1 & 63.8 \\
BioRAG
& 47.6 & 37.1 & {21.4} & 50.3 & 72.9 & 82.2 & 89.2 & 89.0 & 62.0 \\
GeneGPT
& 75.4 & 47.0 & {34.5} & 50.6 & 72.8 & 83.2 & 88.5 & 89.0 & \underline{65.9} \\
\midrule
\rowcolor{blue!12}
\model{}
& \textbf{76.4} & \textbf{54.1} & \textbf{54.6} & \textbf{60.3}
& \textbf{74.9} & \textbf{85.9} & \underline{89.3} & \textbf{89.7}
& \textbf{71.0} \\

\bottomrule
\end{tabular}}
\vspace{2pt}

\begin{minipage}{\textwidth}
\scriptsize
$^{\dagger}$ \textit{w/ Golden Context} is computed only on benchmarks with dataset-provided gold evidence, namely PubMedQA and BioASQ, covering 5,219 examples. ``--'' indicates that gold evidence is unavailable or that the corresponding overall score is not reported.\\
\end{minipage}
\end{table*}

\subsection{Benchmark Datasets}
The benchmark integrates eight publicly available biomedical QA datasets, unified under a consistent JSONL schema and evaluation protocol\footnote{\url{https://huggingface.co/datasets/Shaow/BioHarness_Eval}}.
In total, the benchmark contains 19,302 questions covering seven task types and a broad spectrum of biomedical subdomains, including molecular biology, clinical medicine, and genomics.
Each dataset contributes distinct evaluation signals:
\begin{itemize}
\item Clinical datasets, including MedMCQA~\citep{pal2022medmcqa}, MedQA-US, MedQA-TW, and MedQA-CN~\citep{jin2020medqa}, emphasize diagnostic reasoning and factual recall.
\item Literature-based datasets, including BioASQ~\citep{krithara2023bioasq} and PubMedQA~\citep{jin2019pubmedqa}, assess document-grounded inference and summarization.
\item Genomics-focused datasets, including GeneTuring~\citep{shang2025benchmarking} and SciHorizon-Gene~\citep{qin2025scihorizon,huang2026scihorizon}, stress domain-specific knowledge, entity normalization, and biological reasoning.
\end{itemize}
All datasets are normalized into a unified JSONL format, ensuring consistent access to question text, task type, answer options when applicable, and ground-truth annotations.

\subsection{Baseline Methods}
Following the controlled experimental platform described above, all baselines are evaluated using the same processed corpus, dense retriever, reranker when applicable, answer backbone, constrained decoding strategy, and evaluation pipeline.
The compared methods cover retrieval-free, oracle-evidence, dense retrieval, query augmentation, reranking, reflective retrieval, graph-based retrieval, and biomedical tool-augmented retrieval settings.

\noindent\textbf{Retrieval-free and oracle settings.}
We include \textit{No-Context} as a retrieval-free baseline and \textit{w/ Golden Context} as an oracle-evidence setting using dataset-provided gold evidence when available.

\noindent\textbf{Dense retrieval and query augmentation.}
We compare with \textit{DenseRAG}~\citep{karpukhin2020dense}, and \textit{HyDE}~\citep{gao2023precise}.

\noindent\textbf{Reflective and iterative retrieval.}
We include \textit{DenseRAG}+\textit{Rerank}, \textit{IRCoT}~\citep{trivedi2023ircot}, \textit{Chain-of-Note}~\citep{yu2024chain}, \textit{Self-RAG}~\citep{asai2023self}, and \textit{FLARE}~\citep{zhang2024enhancing}.

\noindent\textbf{Graph-based retrieval.}
We compare with \textit{GraphRAG}~\citep{edge2024local}, \textit{LightRAG}~\citep{guo2024lightrag}, and \textit{PathRAG}~\citep{chen2026pathrag}.

\noindent\textbf{Biomedical and tool-augmented retrieval.}
We further compare with \textit{BiomedRAG}~\citep{li2025biomedrag}, \textit{BioRAG}~\citep{wang2024biorag}, and \textit{GeneGPT}~\citep{jin2024genegpt}.

\section{Results}
\begin{figure*}
\centering
\includegraphics[width=\linewidth]{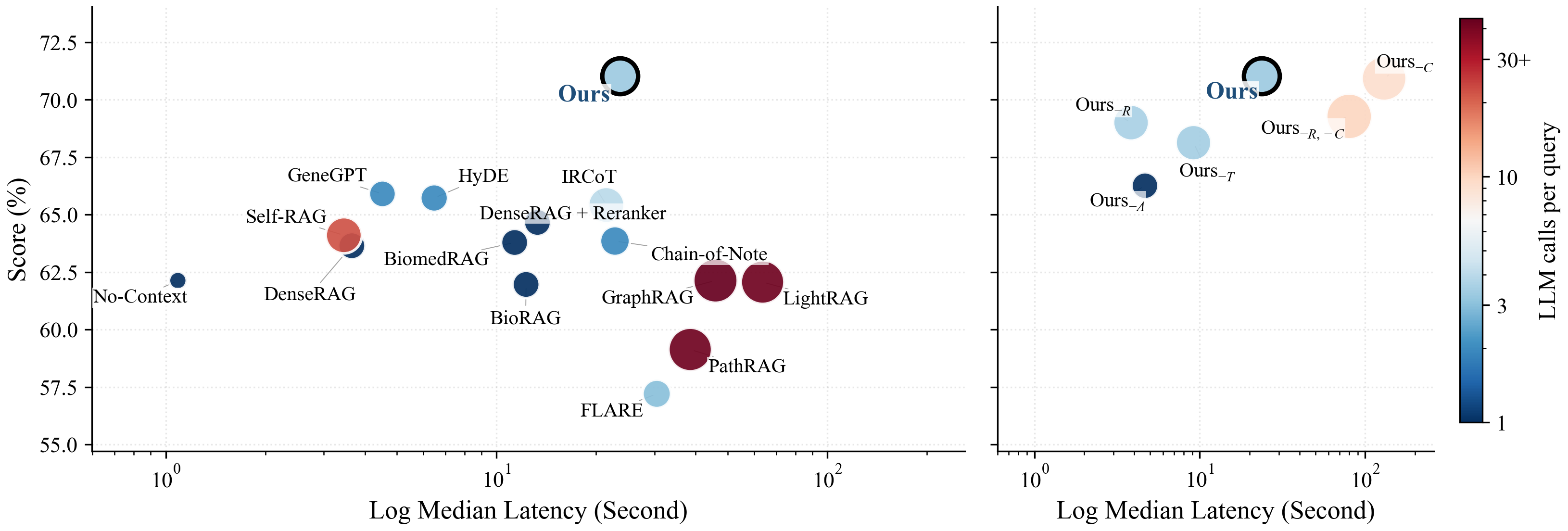}
\caption{
Score--efficiency comparison on the unified 19,302-question benchmark.
The left panel compares \model{} with all baselines, and the right panel shows component ablations of \model{}.
Each point reports the pooled score against median per-question latency on a log-scaled horizontal axis.
Bubble area denotes the average total token count per query, and color denotes the average number of LLM calls per query.
}
\label{fig:efficiency}
\end{figure*}
\subsection{Main Results on the Unified Biomedical QA Benchmark}

Table~\ref{tab:main_results} reports the main comparison on eight biomedical QA benchmarks.
\model{} achieves the best overall performance, improving the strongest non-oracle baseline from 65.9 to 71.0.
It obtains the best result on seven of the eight datasets, with MedQA-TW being the only exception where Self-RAG is slightly higher.
The largest gains appear on benchmarks requiring precise biomedical grounding: \model{} substantially improves GeneTuring and SciHorizon, suggesting that adaptive evidence control is more effective than fixed retrieval or tool-use strategies. 

The comparison also shows that stronger retrieval alone is insufficient.
Dense retrieval, reranking, reflective retrieval, and graph-based retrieval each improve some settings, but their gains remain uneven across the benchmark.
Similarly, biomedical tool-augmented methods such as BioRAG and GeneGPT improve entity-centric tasks, but do not consistently generalize to the full benchmark.
These results suggest that biomedical QA benefits more from adaptive control over retrieval, reranking, structured-resource invocation, and evidence repair than from any single fixed retrieval architecture.

The `w/ Golden Context' provides an evidence upper bound only for PubMedQA and BioASQ: \model{} is close to the oracle on PubMedQA but remains farther from it on BioASQ, reflecting the greater difficulty of multi-format questions that require fine-grained normalization and synthesis across snippets.
In general, the results support the central design of \model{}: an accurate biomedical quality assessment requires not only the retrieval of relevant documents, but also the assembly and repair of evidence according to the type of question, the substrate of the biomedical entity, and the available structured resources.

\begin{figure*}
    \centering
    \includegraphics[width=\linewidth]{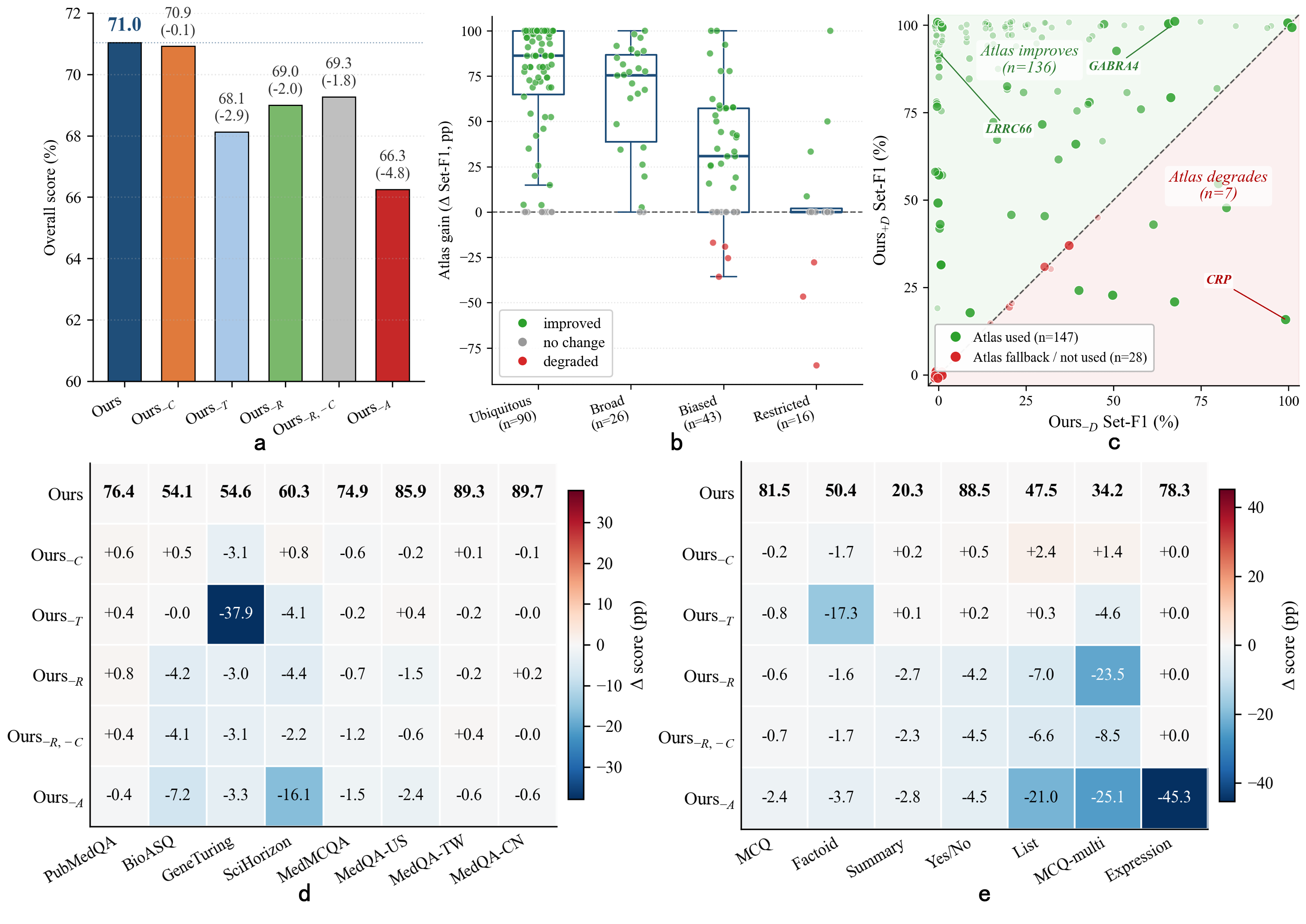}
\caption{
Component ablation and atlas-routed evidence analysis of \model{}.
(a) Pooled overall score under major component ablations on the unified benchmark.
(b) Distribution of atlas-induced Set-F1 changes on atlas-relevant expression questions, grouped by tissue-specificity class.
(c) Paired comparison between \model{} and \model{}$_{-D}$ on atlas-routed cases, where \model{}$_{-D}$ disables atlas-derived measurement access; points above the diagonal indicate cases improved by atlas evidence.
(d) Per-dataset score changes relative to the full model.
(e) Per-task score changes relative to the full model.
Here, \model{}$_{-A}$ disables REPL iterative reasoning escalation while retaining fast-path tool-derived grounding, \model{}$_{-C}$ removes cascade routing and forces all questions into the agent route, \model{}$_{-T}$ disables the Biomedical Tool Layer, \model{}$_{-R}$ removes question-conditioned reranking, and \model{}$_{-D}$ disables atlas-derived measurement access.
}
    \label{fig:ablation}
\end{figure*}

\subsection{Efficiency of Grounded Cascade Control}

Figure~\ref{fig:efficiency} summarizes the score--compute trade-off on the full benchmark.
We measure median end-to-end latency per question, average token usage, and average LLM calls under the same evaluation setting as Table~\ref{tab:main_results}.
For graph-based retrieval methods, latency includes query-time graph construction and reasoning, so the reported operating point reflects a single-shot deployment scenario without cross-query caching.

\model{} lies on the empirical efficient frontier.
It achieves the highest pooled score, improving over the strongest non-oracle baseline from 65.9 to 71.0, while using a moderate compute budget.
In contrast, several methods that spend substantially more computation do not obtain comparable scores.
GraphRAG, PathRAG, and LightRAG require many LLM calls and large token budgets because they construct or traverse query-local graph structures, yet their scores remain below strong dense, query-augmentation, or reflective retrieval baselines.
This indicates that the gain of \model{} is not simply due to spending more inference compute, but to allocating computation according to evidence sufficiency.

The ablation panel further clarifies the role of cascade routing.
Removing cascade control, \model{}$_{-C}$, forces all questions into the agent route and gives nearly the same pooled score as the full pipeline, but shifts markedly toward the high-latency and high-token regime.
The full grounded cascade recovers the same accuracy level with much lower latency and fewer tokens by preserving confident fast-path answers and escalating only uncertain, weakly grounded, or substrate-mismatched cases.
Conversely, disabling REPL iterative reasoning escalation, \model{}${-A}$, yields a cheaper but substantially lower-scoring configuration, confirming that iterative repair is useful when needed but should not be applied indiscriminately.

Overall, the efficiency analysis supports the central design choice of \model{}: literature retrieval, question-conditioned reranking, biomedical tools, atlas access, and REPL-style repair should be treated as a staged evidence-allocation policy rather than as uniformly applied modules.
The cascade preserves the score benefit of iterative repair while avoiding unnecessary agent execution for questions that can already be resolved by the fast path with reranked literature and tool-derived grounding.
Similarly, atlas access is router-gated rather than globally enabled: only a small fraction of benchmark queries are routed to the Atlas Evidence Layer, indicating that structured measurement queries are rare at the benchmark level but require a dedicated substrate when they arise.
This behavior supports the role of \model{} as a substrate-aware harness that selectively allocates computation and switches evidence substrates, instead of applying all retrieval, tool, and atlas operations to every question.

\begin{figure*}
    \centering
    \includegraphics[width=\linewidth]{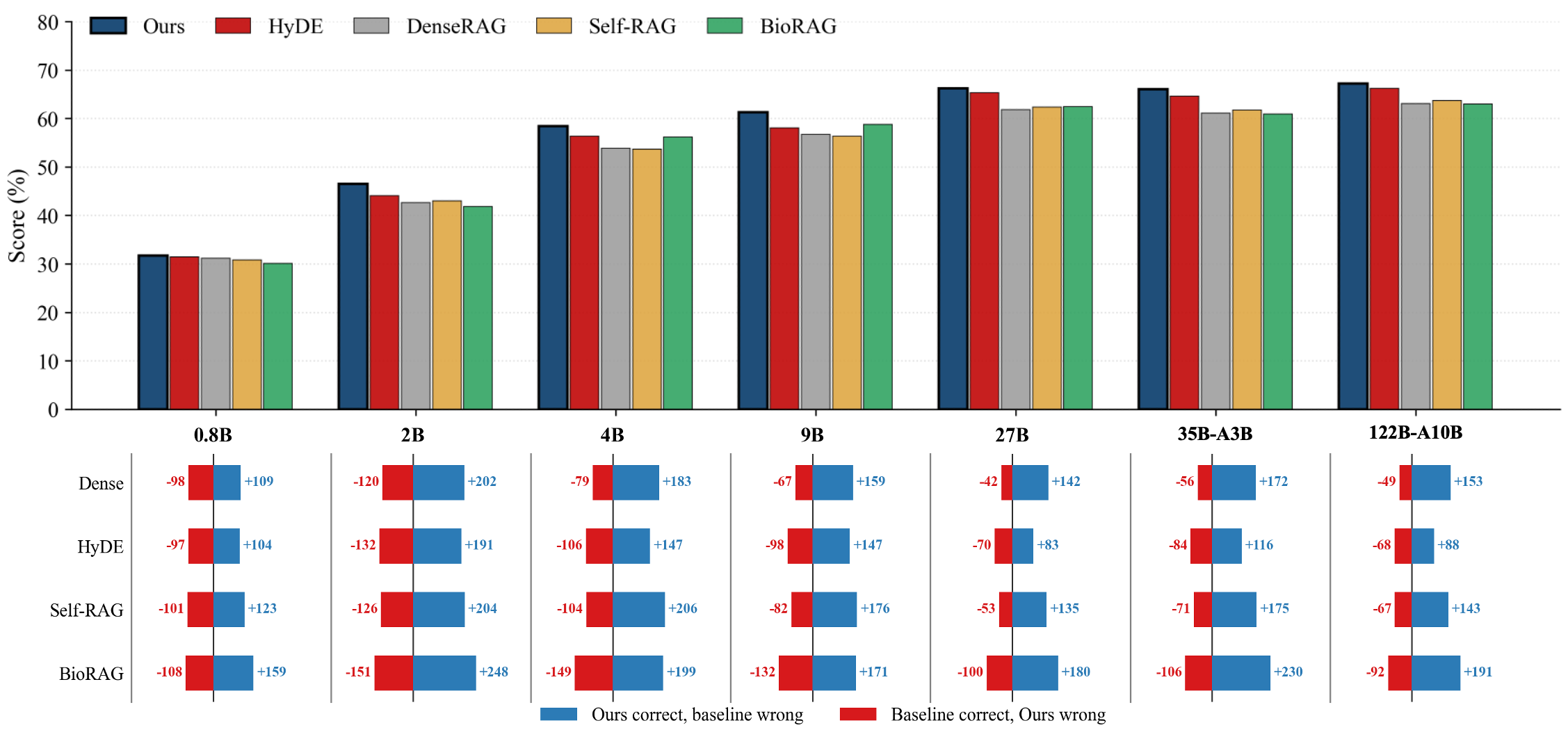}
    \caption{
    Backbone scaling analysis of \model{} on a deterministic 10\% subsample of the unified benchmark.
The top panel reports the pooled score across seven Qwen3.5 backbones ranging from 0.8B to 122B-A10B.
The bottom panel shows paired item-level comparisons between \model{} and representative baselines.
Blue bars denote items where \model{} obtains a higher item-level score than the baseline, and red bars denote the converse.
}
    \label{fig:modelcheck}
\end{figure*}
\subsection{Component Ablation Studies}

Figure~\ref{fig:ablation} analyzes the contributions of the major retrieval, control, repair, and atlas components of \model{}.
Disabling REPL iterative reasoning escalation (\model{}$_{-A}$) yields the largest pooled degradation in the current analysis, reducing the score from 71.0 to 66.3.
Disabling the Biomedical Tool Layer (\model{}$_{-T}$) reduces the score to 68.1, while removing question-conditioned reranking (\model{}$_{-R}$) reduces it to 69.0.
Removing both reranking and cascade control (\model{}$_{-R,-C}$) yields 69.3.
Removing cascade control alone (\model{}$_{-C}$), which forces all questions into the agent route, gives a similar pooled score of 70.9 but changes the error profile.
This indicates that the grounded cascade is not only a cost-saving mechanism; it also protects already-grounded fast-path answers from unnecessary agent-side rewriting or over-processing.

Panels (d) and (e) show where these changes occur.
At the dataset level, disabling the Biomedical Tool Layer causes the largest loss on GeneTuring, consistent with the importance of tool-based alias resolution, canonical entity grounding, and identifier normalization for gene-centric questions.
By contrast, removing cascade control causes only a small pooled change, but it slightly degrades GeneTuring, where many questions can already be answered correctly through fast-path retrieval plus tool-derived grounding.
This supports the intended role of the grounded gate: it escalates uncertain or substrate-mismatched cases while preserving confident fast-path answers when additional agent reasoning is more likely to perturb than repair the answer.
Disabling REPL iterative escalation mainly affects SciHorizon and BioASQ, indicating that iterative repair is most useful when the model must recover or synthesize evidence beyond a single retrieved passage.
At the task level, this degradation is most pronounced for expression, multi-answer choice, and list-style questions, which often require evidence aggregation or repeated checking.
Removing reranking also hurts BioASQ and SciHorizon, and particularly affects list and multi-answer settings, supporting its role in selecting answer-bearing evidence from topically similar biomedical abstracts.

Panels (b) and (c) isolate the Atlas Evidence Layer, which is router-gated and therefore not well summarized by a single pooled benchmark score.
Panel (b) shows that atlas-derived measurements provide substantial gains for many atlas-relevant expression questions, especially when tissue-level expression patterns are broad or under-represented in the retrieved literature.
Panel (c) further compares \model{} with \model$_{-D}$ on atlas-routed cases.
Most points lie above the diagonal, indicating that atlas evidence improves the routed examples, while degradation is rare and concentrated in a small number of cases.

Qualitatively, the paired atlas analysis reveals two dominant improvement modes and one residual failure mode.
For genes with under-reported literature evidence, atlas measurements recover missing expression sites, as in \texttt{GABRA4}, where the literature-only prediction retrieves only brain while atlas evidence completes the ground truth with cardiac expression.
For sparsely characterized genes, atlas evidence can be decisive: \texttt{LRRC66} receives no usable tissue assignment from literature retrieval, whereas atlas-derived measurements recover a gastrointestinal expression pattern with high Set-F1.
The main degradation mode occurs when the literature already provides a precise restricted answer, but the atlas introduces broad low-level off-target signals, as in \texttt{CRP}, where over-trusting noisy atlas breadth expands the answer beyond liver and reduces Set-F1.
Thus, atlas access is most useful when literature priors are incomplete or weak, but requires conservative filtering when structured measurements are broader than the benchmark target.

Overall, the ablation results support the substrate-aware design of \model{}.
Literature reranking improves answer-bearing evidence selection, biomedical tools repair entity-level grounding failures, REPL iterative escalation repairs unresolved cases beyond the fast path, and atlas-derived measurements supply a targeted substrate for measurement-centric questions.

\subsection{Backbone Scaling Analysis}

To examine whether the gains of \model{} depend on a particular answer-model scale, we re-evaluated \model{} and four representative backbone-swappable baselines, DenseRAG, HyDE, Self-RAG, and BioRAG, on a deterministic 10\% subsample of the unified benchmark in Figure~\ref{fig:modelcheck}.

As shown in the top panel, \model{} obtains the highest pooled score at every tested backbone size.
The margin is small at 0.8B, where all methods remain in a low-score regime, but becomes clearer from 2B onward.
Although larger backbones improve all methods, increasing answer-model capacity alone does not remove the gap between \model{} and the representative retrieval baselines included in this analysis.

The bottom panel provides a paired item-level view of the same trend.
Across all tested backbones and all four baselines, \model{} maintains a positive win--loss balance, with more items improved by \model{} than items where the corresponding baseline performs better.
This indicates that the subsample-level gain is not driven only by a small number of high-scoring outliers, but reflects consistent item-level improvements.

Together with the results in Table~\ref{tab:main_results}, this analysis supports the claim that the gains of \model{} do not arise merely from using a particular answer backbone.
Rather, they are consistent with the intended design of the harness: improving the evidence available before generation through literature evidence ranking, biomedical entity grounding, and executable evidence repair.

\begin{figure}[t]
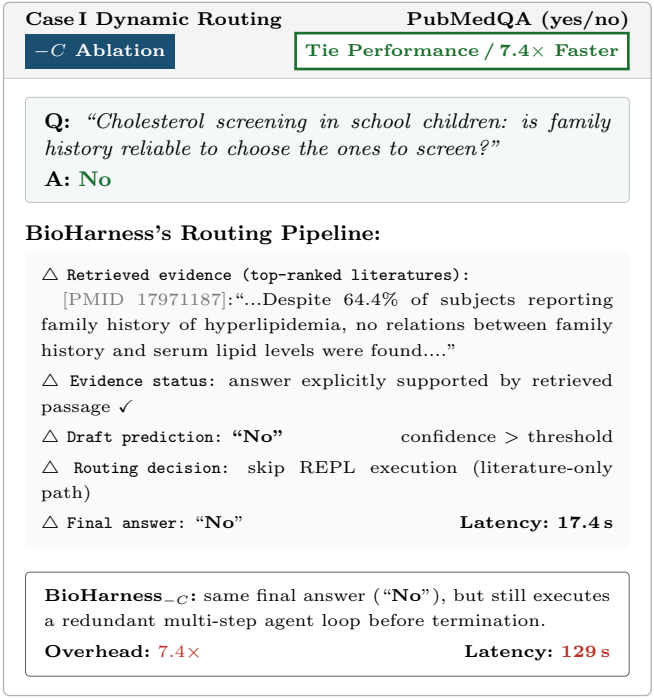

\centering
\begin{tcolorbox}[enhanced, width=\linewidth, colback=white, colframe=cCardFrame,
  boxrule=0.5pt, arc=2pt, left=5pt, right=5pt, top=4pt, bottom=4pt,
  fonttitle=\bfseries\small, coltitle=black,
  colbacktitle=cCardTitleBg, titlerule=0.3pt,
  title={Case\,I\ \textbf{Dynamic Routing}\quad
         \textsc{\hfill PubMedQA (yes/no)}\\
         \ablBadge{cAbC}{$-C$ Ablation}\hfill \verdictBadge{cWin}{Tie Performance\,/\,7.4$\times$ Faster}}]
\footnotesize
\begin{tcolorbox}[
  colback=cBlockBg,
  colframe=cCardFrame,
  boxrule=0.3pt,
  arc=1.5pt,
  left=4pt,right=4pt,top=3pt,bottom=3pt]
\footnotesize
\textbf{Q:}\ 
\emph{``Cholesterol screening in school children: is family history
reliable to choose the ones to screen?''}

\vspace{2pt}

\textbf{A:}\ 
{\color{cWin}\bfseries No}
\end{tcolorbox}

\par\vspace{2pt}

\textbf{\model's Routing Pipeline:}

\par\vspace{2pt}

\colorbox{cCodeBg}{%
\parbox{\dimexpr\linewidth-2\fboxsep}{\scriptsize

$\triangle$ \texttt{Retrieved evidence (top-ranked literatures):}

\quad \textcolor{cMuted}{[PMID 17971187]}:``...Despite 64.4\% of subjects reporting family history of hyperlipidemia, no relations between family history and serum lipid levels were found....''

\vspace{2pt}

$\triangle$ \texttt{Evidence status:} answer explicitly supported by retrieved passage \checkmark

\vspace{2pt}

$\triangle$ \texttt{Draft prediction:} \textbf{``No''}  \hfill confidence $>$ threshold

\vspace{2pt}

$\triangle$ \texttt{Routing decision:}
skip REPL execution (literature-only path)

\vspace{2pt}

$\triangle$ \texttt{Final answer:} ``\textbf{No}''
\hfill \textbf{Latency: 17.4\,s}
}}

\par\vspace{4pt}

\begin{tcolorbox}[
  colback=white,
  colframe=cTie,
  boxrule=0.35pt,
  arc=1.5pt,
  left=4pt,right=4pt,top=3pt,bottom=3pt]
\scriptsize
\textbf{\model$_{-C}$:} same final answer (``\textbf{No}''), but still
executes a redundant multi-step agent loop before termination.

\vspace{2pt}
\textbf{Overhead:} {\color{cLoss}\bfseries 
$7.4\times$}\hfill
\textbf{Latency:} {\color{cLoss}\bfseries 129\,s}

\end{tcolorbox}
\end{tcolorbox}
\caption{Cascade routing avoids unnecessary REPL-style evidence repair.}
\label{fig:case_cholesterol}
\end{figure}


\begin{figure}[t]
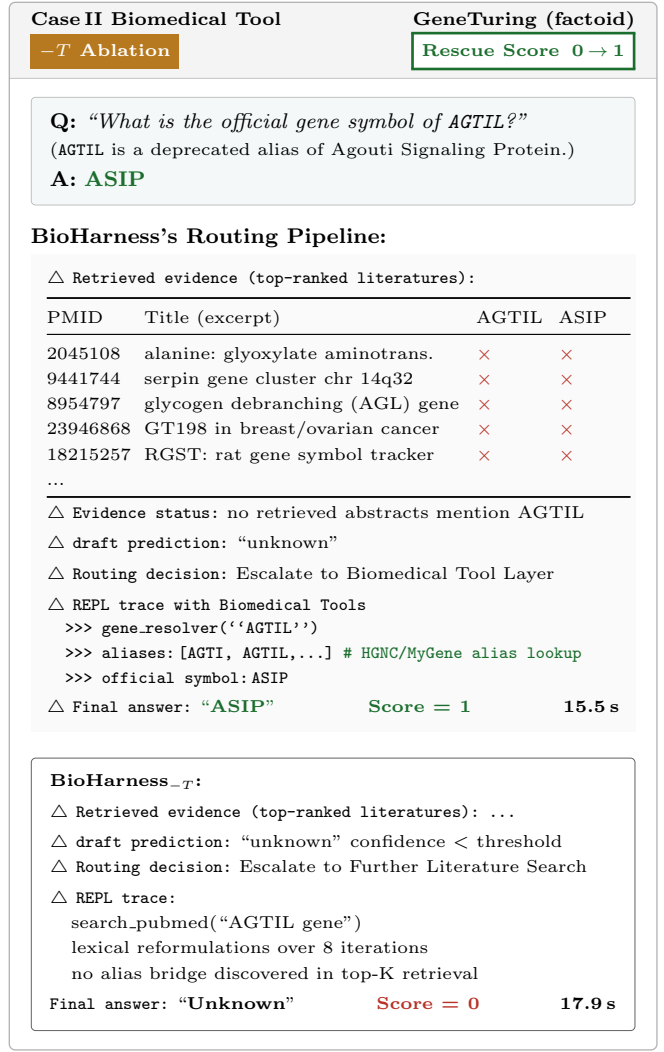

\centering
\begin{tcolorbox}[
  enhanced,
  width=\linewidth,
  colback=white,
  colframe=cCardFrame,
  boxrule=0.5pt,
  arc=2pt,
  left=5pt,right=5pt,top=4pt,bottom=4pt,
  fonttitle=\bfseries\small,
  coltitle=black,
  colbacktitle=cCardTitleBg,
  titlerule=0.3pt,
  title={
    Case\,II\ \textbf{Biomedical Tool}\hfill
    \textsc{GeneTuring (factoid)}\\
    \ablBadge{cAbT}{$-T$ Ablation}\hfill
    \verdictBadge{cWin}{Rescue Score \,0\,$\to$\,1}
  }]

\footnotesize


\begin{tcolorbox}[
  colback=cBlockBg,
  colframe=cCardFrame,
  boxrule=0.3pt,
  arc=1.5pt,
  left=4pt,right=4pt,top=3pt,bottom=3pt]
\footnotesize

\textbf{Q:}
\emph{``What is the official gene symbol of \texttt{AGTIL}?''} 

{\scriptsize (\texttt{AGTIL} is a deprecated alias of Agouti Signaling Protein.)}

\vspace{2pt}

\textbf{A:}
{\color{cWin} \textbf{ASIP}} 

\vspace{1pt}

\end{tcolorbox}

\par\vspace{2pt}

\textbf{\model's Routing Pipeline:}

\par\vspace{2pt}

\colorbox{cCodeBg}{%
\parbox{\dimexpr\linewidth-2\fboxsep}{\scriptsize
$\triangle$ \texttt{Retrieved evidence (top-ranked literatures):}
\par\smallskip

{\scriptsize
\setlength{\tabcolsep}{2pt}
\begin{tabular}{@{}p{1.15cm}p{4.25cm}p{0.95cm}p{0.95cm}@{}}
\toprule PMID & Title (excerpt) & AGTIL & ASIP \\
\midrule
2045108  & alanine: glyoxylate aminotrans.  &
\textcolor{cLoss}{$\times$} &
\textcolor{cLoss}{$\times$} \\

9441744  & serpin gene cluster chr 14q32 &
\textcolor{cLoss}{$\times$} &
\textcolor{cLoss}{$\times$} \\

8954797  & glycogen debranching (AGL) gene &
\textcolor{cLoss}{$\times$} &
\textcolor{cLoss}{$\times$} \\

23946868 & GT198 in breast/ovarian cancer &
\textcolor{cLoss}{$\times$} &
\textcolor{cLoss}{$\times$} \\

18215257 & RGST: rat gene symbol tracker &
\textcolor{cLoss}{$\times$} &
\textcolor{cLoss}{$\times$} \\
...\\
\bottomrule
\end{tabular}

\vspace{2pt}

$\triangle$ \texttt{Evidence status:} no retrieved abstracts mention AGTIL

\vspace{2pt}

$\triangle$ \texttt{draft prediction:}
``unknown''

\vspace{2pt}

$\triangle$ \texttt{Routing decision:}
Escalate to Biomedical Tool Layer

\vspace{2pt}

$\triangle$ \texttt{REPL trace with Biomedical Tools}
\vspace{2pt}

\parbox{\dimexpr\linewidth-2\fboxsep}{\scriptsize\ttfamily
\quad >>> gene\_resolver(``AGTIL'')

\quad >>> aliases:\,[AGTI, AGTIL,...] \textcolor{cWin}{\# HGNC/MyGene alias lookup}

\quad >>> official symbol:\,{ASIP}
}
\vspace{4pt}

$\triangle$ \texttt{Final answer:}
\textcolor{cWin}{``\textbf{ASIP}''}
\hfill
\textcolor{cWin}{\bfseries Score = 1}
\hfill
\textbf{15.5\,s}
}}}

\par\smallskip

\scriptsize

\par\vspace{2pt}

\begin{tcolorbox}[
  colback=white,
  colframe=cTie,
  boxrule=0.35pt,
  arc=1.5pt,
  left=4pt,right=4pt,top=3pt,bottom=3pt]
\scriptsize
\textbf{\model$_{-T}$: }

\vspace{2pt}

$\triangle$ \texttt{Retrieved evidence (top-ranked literatures): ...}

\vspace{2pt}

$\triangle$ \texttt{draft prediction:}
``unknown'' confidence $<$ threshold

$\triangle$ \texttt{Routing decision:}
Escalate to Further Literature Search

\vspace{2pt}

$\triangle$ \texttt{REPL trace:}

\quad search\_pubmed(``AGTIL gene'')

\quad lexical reformulations over 8 iterations

\quad no alias bridge discovered in top-K retrieval

\vspace{2pt}

\texttt{Final answer:}
``\textbf{Unknown}''
\hfill
\textcolor{cLoss}{\bfseries Score = 0}
\hfill
\textbf{17.9\,s}

\vspace{2pt}
\end{tcolorbox}







\end{tcolorbox}

\caption{Biomedical Tool Layer resolves alias chains that literature retrieval cannot bridge.}
\label{fig:case_agtil}
\end{figure}
\begin{figure*}[t]
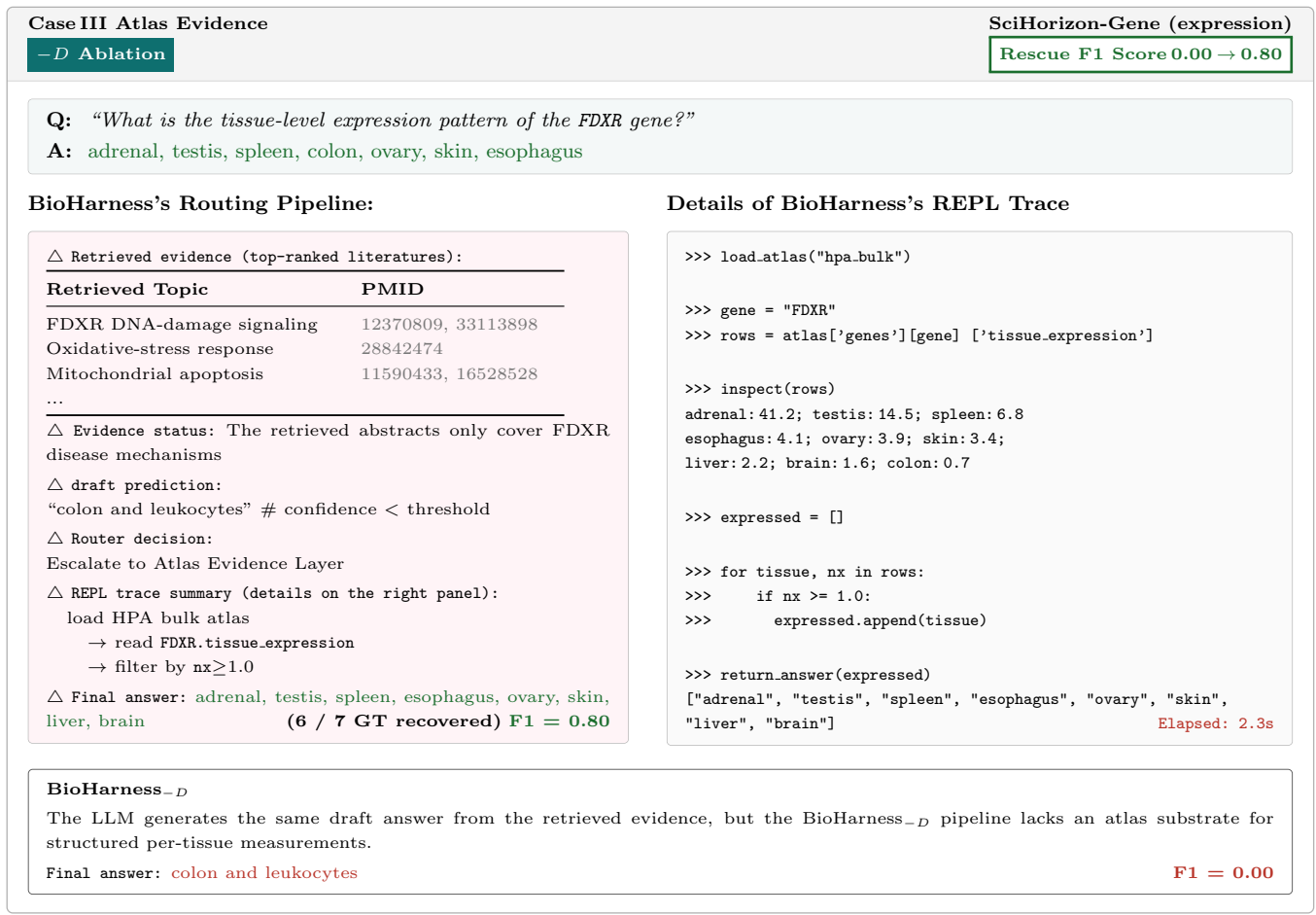

\centering

\begin{tcolorbox}[
  enhanced,
  width=\linewidth,
  colback=white,
  colframe=cCardFrame,
  boxrule=0.5pt,
  arc=2pt,
  left=5pt,right=5pt,top=4pt,bottom=4pt,
  fonttitle=\bfseries\small,
  coltitle=black,
  colbacktitle=cCardTitleBg,
  titlerule=0.2pt,
  title={
    Case\,III\ \textbf{Atlas Evidence}\hfill
    \textsc{SciHorizon-Gene (expression)}\\
    \ablBadge{cAbD}{$-D$ Ablation}\hfill
    \verdictBadge{cWin}{Rescue F1 Score\,0.00\,$\to$\,0.80}
  }]

\footnotesize


\begin{tcolorbox}[
  colback=cBlockBg,
  colframe=cCardFrame,
  boxrule=0.2pt,
  arc=1.5pt,
  left=4pt,right=4pt,top=2pt,bottom=2pt]

\footnotesize

\textbf{Q: }
\emph{``What is the tissue-level expression pattern of the \texttt{FDXR} gene?''}

\vspace{2pt}

\textbf{A: }
{\color{cWin}
adrenal, testis, spleen, colon, ovary, skin, esophagus
}

\end{tcolorbox}

\par\vspace{2pt}


\noindent
\begin{minipage}[t]{0.475\linewidth}



\textbf{\model's Routing Pipeline:}


\begin{tcolorbox}[
  colback=cRouteBg,
  colframe=cCardFrame,
  boxrule=0.25pt,
  arc=1.5pt,
  left=4pt,right=4pt,top=2pt,bottom=2pt]

\scriptsize

\vspace{2pt}

$\triangle$ \texttt{Retrieved evidence (top-ranked literatures):}
\vspace{1pt}

{\scriptsize
\setlength{\tabcolsep}{2pt}
\begin{tabular}{@{}p{0.54\linewidth}p{0.36\linewidth}@{}}
\toprule
\textbf{Retrieved Topic} & \textbf{PMID} \\
\midrule
FDXR DNA-damage signaling
&
\textcolor{cMuted}{12370809, 33113898}
\\

Oxidative-stress response
&
\textcolor{cMuted}{28842474}
\\

Mitochondrial apoptosis
&
\textcolor{cMuted}{11590433, 16528528}
\\
...
\\
\bottomrule
\end{tabular}}

\vspace{2pt}

$\triangle$ \texttt{Evidence status:} The retrieved abstracts only cover FDXR disease mechanisms

\vspace{2pt}

$\triangle$ \texttt{draft prediction:}

``colon and leukocytes'' \# confidence $<$ threshold

\vspace{2pt}

$\triangle$ \texttt{Router decision:}

Escalate to Atlas Evidence Layer

\vspace{2pt}

$\triangle$ \texttt{REPL trace summary (details on the right panel):}

\quad load HPA bulk atlas

\qquad $\rightarrow$ read \texttt{FDXR.tissue\_expression}

\qquad $\rightarrow$ filter by \texttt{nx}$\geq$1.0

\vspace{2pt}

$\triangle$ \texttt{Final answer:} {\color{cWin} adrenal, testis, spleen, esophagus, ovary, skin, liver, brain}
\hfill
\textbf{(6 / 7 GT recovered)} \textcolor{cWin}{\bfseries F1 = 0.80} 

\end{tcolorbox}
\end{minipage}
\hfill
\begin{minipage}[t]{0.495\linewidth}



\textbf{Details of \model's REPL Trace}


\begin{tcolorbox}[
  colback=cCodeBg,
  colframe=cCardFrame,
  boxrule=0.25pt,
  arc=1.5pt,
  left=4pt,right=4pt,top=2pt,bottom=2pt]

\scriptsize\ttfamily
\vspace{2pt}

>>> load\_atlas("hpa\_bulk")\\[2pt]

>>> gene = "FDXR"\\
>>> rows = atlas['genes'][gene]  ['tissue\_expression']\\[2pt]

>>> inspect(rows)\\
\qquad adrenal:\,41.2; testis:\,14.5; spleen:\,6.8 \\
\qquad esophagus:\,4.1; ovary:\,3.9; skin:\,3.4; \\ \qquad liver:\,2.2; brain:\,1.6; colon:\,0.7\\[2pt]

>>> expressed = []\\[2pt]

>>> for tissue, nx in rows:\\
>>> \qquad if nx >= 1.0:\\
>>> \qquad\quad expressed.append(tissue)\\[2pt]

>>> return\_answer(expressed)\\
\qquad ["adrenal", "testis", "spleen", "esophagus", "ovary", "skin", "liver", "brain"] \hfill {\color{cLoss}Elapsed: 2.3s}
\end{tcolorbox}
\end{minipage}
    
\par\vspace{4pt}

\begin{tcolorbox}[
  colback=white,
  colframe=cTie,
  boxrule=0.35pt,
  arc=1.5pt,
  left=4pt,right=4pt,top=2pt,bottom=2pt]

\scriptsize

\textbf{\model$_{-D}$}

\vspace{2pt}

The LLM generates the same draft answer from the retrieved evidence, but the \model$_{-D}$ pipeline lacks an atlas substrate for structured per-tissue measurements.

\vspace{2pt}

\texttt{Final answer: }{\color{cLoss} colon and leukocytes}
\hfill
\textcolor{cLoss}{\bfseries F1 = 0.00}

\end{tcolorbox}
\end{tcolorbox}

\caption{Atlas-derived evidence provides structured tissue-level measurement grounding absent from the retrieved literature.}

\label{fig:case_fdxr}

\end{figure*}








\subsection{Mechanistic Analysis of Evidence Assembly}

\paragraph{\textbf{Case I: Retrieval Saturation and Dynamic Escalation.}}

The case study from Figure~\ref{fig:case_cholesterol} illustrates the primary role of the cascade controller: avoiding unnecessary escalation when the retrieved literature already provides explicit answer-supporting evidence.
The top-ranked literature passage directly states that family history was not associated with serum lipid levels, allowing \model{} to produce a grounded negative answer from the literature-only path.
The controller therefore terminates after the confidence and grounding checks are satisfied.
In contrast, the $-C$ ablation disables selective routing by forcing REPL execution.
Although the retrieved evidence is already sufficient, the system still executes a redundant multi-step repair loop before stopping.
This additional trajectory does not improve the final prediction, which remains ``No'', but increases inference overhead from 17.4,s to 129,s.
Rather than invoking REPL-style repair for every query, \model{} escalates only when the retrieved evidence remains ambiguous, incomplete, or weakly grounded. 

\paragraph{\textbf{Case II: Entity Grounding via Biomedical Tool Calls.}}

The case study in Figure~\ref{fig:case_agtil} illustrates a representative biomedical entity-grounding failure that cannot be resolved through literature retrieval alone.
The query uses \texttt{AGTIL}, a deprecated alias of the canonical gene symbol \texttt{ASIP}.
Because this alias is not represented in the retrieved literature context, dense retrieval remains trapped in an incorrect lexical neighborhood, returning unrelated genes and symbol fragments despite repeated retrieval attempts.
The failure is therefore not due to insufficient retrieval depth, but to the absence of a deterministic alias bridge between the surface mention and the canonical biomedical entity.
Even iterative literature reformulation in \model$_{-T}$ remains confined to lexical search and eventually terminates with ``unknown'' after repeated unsuccessful retrieval rounds.

In contrast, the full \model{} pipeline identifies the lack of grounded evidence and escalates to REPL-style evidence repair, where the Biomedical Tool Layer resolves the alias through a gene-normalization call.
Rather than continuing open-ended retrieval, the system converts the problem into a structured identifier-resolution task.
The resolver maps the deprecated alias \texttt{AGTIL} to the official HGNC symbol \texttt{ASIP}, immediately recovering the correct answer.
This case demonstrates that biomedical QA errors can originate from entity-normalization failures rather than reasoning deficiencies.
For such questions, additional literature retrieval provides diminishing returns because the missing link is structural rather than textual.
The Biomedical Tool Layer therefore acts as an entity-grounding substrate that bridges aliases, deprecated symbols, and identifier inconsistencies that are difficult to resolve with retrieval-only pipelines.

\paragraph{\textbf{Case III: Atlas-Driven Evidence beyond Retrieval.}}

Figure~\ref{fig:case_fdxr} illustrates a representative failure mode where the required evidence substrate is structured rather than textual.
Although literature retrieval identifies passages related to \texttt{FDXR}, the retrieved abstracts primarily discuss DNA-damage signaling, oxidative stress, and mitochondrial apoptosis rather than tissue-level expression measurements.
As a result, the literature-conditioned draft prediction remains weakly grounded and incorrectly infers ``colon and leukocytes'' from indirect contextual associations.

The cascade controller detects both the low-confidence prediction and the mismatch between the retrieved evidence type and the question requirement.
Rather than continuing iterative literature retrieval, \model{} escalates to the Atlas Evidence Layer, which exposes structured tissue-level measurements through an executable atlas interface.
As shown in Figure~\ref{fig:case_fdxr}, the REPL trace loads Human Protein Atlas tissue-expression records for \texttt{FDXR}, applies a deterministic atlas threshold, and returns an atlas-derived tissue set from the structured substrate.

This case highlights an important limitation of retrieval-only biomedical QA systems: some biomedical questions are fundamentally not literature-summarization problems.
Expression profiles, atlas measurements, and structured omics annotations often exist as database-native signals rather than declarative textual statements in abstracts.
The Atlas Evidence Layer enables \model{} to switch from unstructured literature reasoning to executable biological measurement grounding, improving the item-level score on measurement-centric biomedical QA.

\section{Discussion}
This work highlights evidence-substrate mismatch as a key failure mode in biomedical RAG.
\model{} addresses this issue by coordinating literature retrieval, biomedical entity tools, and atlas-derived structured measurements within a staged evidence assembly framework.
The results indicate that selective escalation is more effective than indiscriminately invoking multi-step reasoning, because additional execution is useful mainly when it repairs weakly grounded, uncertain, or substrate-mismatched evidence.
This suggests that future biomedical QA systems should be designed as substrate-aware evidence harnesses that connect textual evidence with curated biomedical knowledge resources and biological atlases, rather than treating all evidence as retrievable text.
Future work could improve evidence-sufficiency estimation, expand coverage of structured biomedical resources, and provide clearer provenance for answers supported by evidence from the literature, knowledge-base records, and atlas-derived measurements.

\section{Availability}
\begin{figure*}
    \centering
    \includegraphics[width=\linewidth]{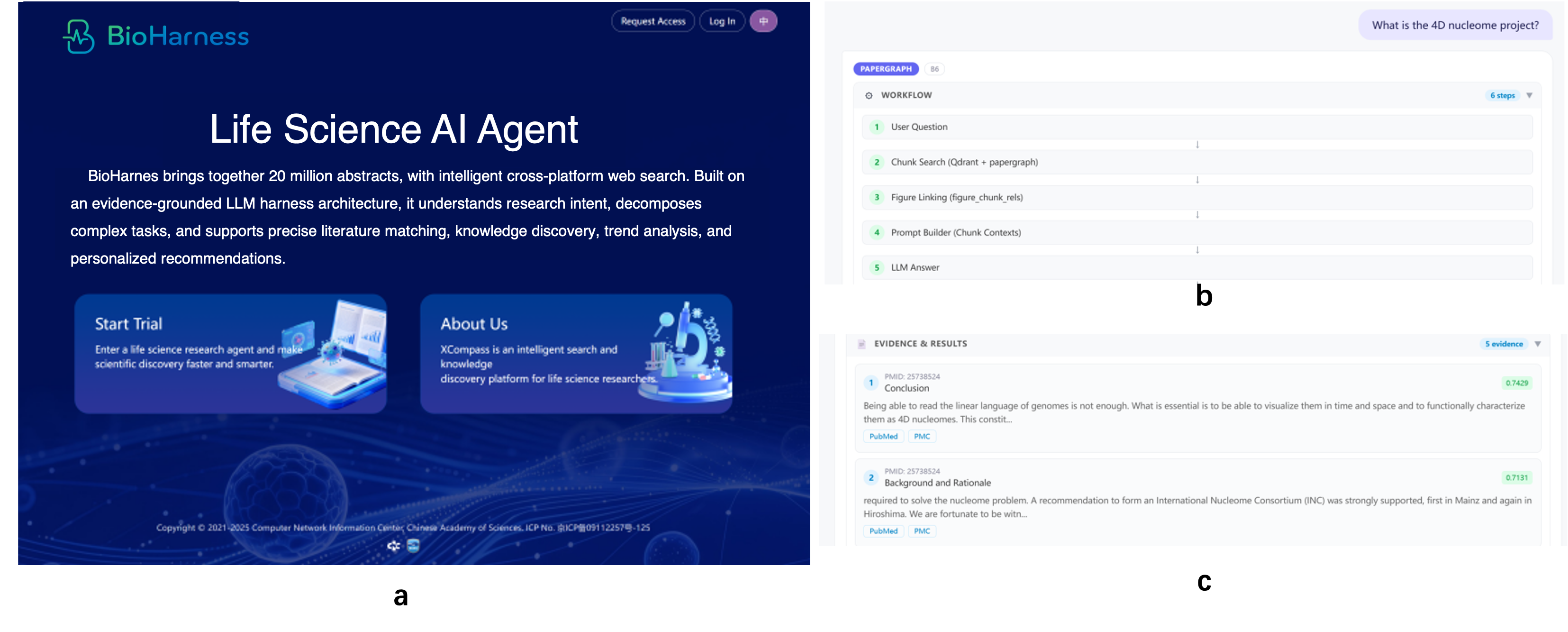}
\caption{
Demo system of \model{}. (a) The BioHarness landing page provides access to the life science question-answering interface. (b) The workflow panel visualizes the execution trace of a user query, including query intake, literature chunk search, figure or document linking, prompt construction, and answer generation.  (c) The evidence panel displays retrieved literature snippets with document identifiers, source links, and relevance scores, enabling users to inspect the evidence associated with the generated answer. 
}
\label{fig:system}
\end{figure*}

We also developed a web-based demo system to illustrate how \model{} can be used in an interactive biomedical QA setting.
As shown in Figure~\ref{fig:system}, the demo exposes both the user-facing question-answering interface and the internal evidence workflow.
For each query, the system records major execution steps such as literature chunk search, figure linking, prompt construction, and answer generation.
The retrieved evidence panel further displays document identifiers, source links, and relevance scores, allowing users to inspect the supporting literature behind the generated answer. 
The purpose of the demo is not to introduce an additional benchmark, but to demonstrate the practical instantiation of the proposed LLM harness.
A public release is planned following stability and security testing.

\section{Acknowledgement}
This work is partially supported by the National Natural Science Foundation of China (No.62506351), the Strategic Priority Research Program of the Chinese Academy of Sciences (No.XDB1350102), the National Natural Science Foundation of China (No.92470204), and the Beijing Natural Science Foundation (No.4254089).
The development of this project was supported by the SciMatrix platform and SciHorizon platform.


\balance
\bibliographystyle{plainnat}
\bibliography{ref,bio}

\begin{thebibliography}{39}
\providecommand{\natexlab}[1]{#1}
\providecommand{\url}[1]{\texttt{#1}}
\expandafter\ifx\csname urlstyle\endcsname\relax
  \providecommand{\doi}[1]{doi: #1}\else
  \providecommand{\doi}{doi: \begingroup \urlstyle{rm}\Url}\fi

\bibitem[Asai et~al.(2023)Asai, Wu, Wang, Sil, and Hajishirzi]{asai2023self}
Akari Asai, Zeqiu Wu, Yizhong Wang, Avirup Sil, and Hannaneh Hajishirzi.
\newblock Self-rag: Learning to retrieve, generate, and critique through self-reflection.
\newblock In \emph{The Twelfth International Conference on Learning Representations}, 2023.

\bibitem[Chen et~al.(2026)Chen, Guo, Yang, Chen, Chen, Liu, Shi, and Yang]{chen2026pathrag}
Boyu Chen, Zirui Guo, Zidan Yang, Yuluo Chen, Junze Chen, Zhenghao Liu, Chuan Shi, and Cheng Yang.
\newblock Pathrag: Pruning graph-based retrieval augmented generation with relational paths.
\newblock In \emph{Proceedings of the AAAI Conference on Artificial Intelligence}, volume~40, pages 30183--30191, 2026.

\bibitem[Consortium et~al.(2022)Consortium, Jones, Karkanias, Krasnow, Pisco, Quake, Salzman, Yosef, Bulthaup, Brown, et~al.]{the2022tabula}
The Tabula~Sapiens Consortium, Robert~C Jones, Jim Karkanias, Mark~A Krasnow, Angela~Oliveira Pisco, Stephen~R Quake, Julia Salzman, Nir Yosef, Bryan Bulthaup, Phillip Brown, et~al.
\newblock The tabula sapiens: A multiple-organ, single-cell transcriptomic atlas of humans.
\newblock \emph{Science}, 376\penalty0 (6594):\penalty0 eabl4896, 2022.

\bibitem[Edge et~al.(2024)Edge, Trinh, Cheng, Bradley, Chao, Mody, Truitt, Metropolitansky, Ness, and Larson]{edge2024local}
Darren Edge, Ha~Trinh, Newman Cheng, Joshua Bradley, Alex Chao, Apurva Mody, Steven Truitt, Dasha Metropolitansky, Robert~Osazuwa Ness, and Jonathan Larson.
\newblock From local to global: A graph rag approach to query-focused summarization.
\newblock \emph{arXiv preprint arXiv:2404.16130}, 2024.

\bibitem[Gao et~al.(2023)Gao, Ma, Lin, and Callan]{gao2023precise}
Luyu Gao, Xueguang Ma, Jimmy Lin, and Jamie Callan.
\newblock Precise zero-shot dense retrieval without relevance labels.
\newblock In \emph{Proceedings of the 61st Annual Meeting of the Association for Computational Linguistics (Volume 1: Long Papers)}, pages 1762--1777, 2023.

\bibitem[George et~al.(2024)George, Fexova, Fuentes, Madrigal, Bi, Iqbal, Kumbham, Nolte, Zhao, Thanki, et~al.]{george2024expression}
Nancy George, Silvie Fexova, Alfonso~Munoz Fuentes, Pedro Madrigal, Yalan Bi, Haider Iqbal, Upendra Kumbham, Nadja~Francesca Nolte, Lingyun Zhao, Anil~S Thanki, et~al.
\newblock Expression atlas update: insights from sequencing data at both bulk and single cell level.
\newblock \emph{Nucleic Acids Research}, 52\penalty0 (D1):\penalty0 D107--D114, 2024.

\bibitem[Guo et~al.(2024)Guo, Xia, Yu, Ao, and Huang]{guo2024lightrag}
Zirui Guo, Lianghao Xia, Yanhua Yu, Tian Ao, and Chao Huang.
\newblock Lightrag: Simple and fast retrieval-augmented generation.
\newblock \emph{arXiv preprint arXiv:2410.05779}, 2\penalty0 (3), 2024.

\bibitem[Gyori et~al.(2022)Gyori, Hoyt, and Steppi]{gyori2022gilda}
Benjamin~M Gyori, Charles~Tapley Hoyt, and Albert Steppi.
\newblock Gilda: biomedical entity text normalization with machine-learned disambiguation as a service.
\newblock \emph{Bioinformatics advances}, 2\penalty0 (1):\penalty0 vbac034, 2022.

\bibitem[Huang et~al.(2026)Huang, Xiao, Qin, Long, Chen, Zhou, and Zhu]{huang2026scihorizon}
Xiaohan Huang, Meng Xiao, Chuan Qin, Qingqing Long, Jinmiao Chen, Yuanchun Zhou, and Hengshu Zhu.
\newblock Scihorizon-gene: Benchmarking llm for life sciences inference from gene knowledge to functional understanding.
\newblock \emph{arXiv preprint arXiv:2601.12805}, 2026.

\bibitem[Jin et~al.(2020)Jin, Pan, Oufattole, Weng, Fang, and Szolovits]{jin2020medqa}
Di~Jin, Eileen Pan, Nassim Oufattole, Wei-Hung Weng, Hanyi Fang, and Peter Szolovits.
\newblock What disease does this patient have? a large-scale open domain question answering dataset from medical exams.
\newblock \emph{arXiv preprint arXiv:2009.13081}, 2020.

\bibitem[Jin et~al.(2019)Jin, Dhingra, Liu, Cohen, and Lu]{jin2019pubmedqa}
Qiao Jin, Bhuwan Dhingra, Zhengping Liu, William Cohen, and Xinghua Lu.
\newblock {P}ub{M}ed{QA}: A dataset for biomedical research question answering.
\newblock In Kentaro Inui, Jing Jiang, Vincent Ng, and Xiaojun Wan, editors, \emph{Proceedings of the 2019 Conference on Empirical Methods in Natural Language Processing and the 9th International Joint Conference on Natural Language Processing (EMNLP-IJCNLP)}, pages 2567--2577, Hong Kong, China, November 2019. Association for Computational Linguistics.
\newblock \doi{10.18653/v1/D19-1259}.
\newblock URL \url{https://aclanthology.org/D19-1259/}.

\bibitem[Jin et~al.(2022)Jin, Yuan, Xiong, Yu, Ying, Tan, Chen, Huang, Liu, and Yu]{jin2022biomedical}
Qiao Jin, Zheng Yuan, Guangzhi Xiong, Qianlan Yu, Huaiyuan Ying, Chuanqi Tan, Mosha Chen, Songfang Huang, Xiaozhong Liu, and Sheng Yu.
\newblock Biomedical question answering: a survey of approaches and challenges.
\newblock \emph{ACM Computing Surveys (CSUR)}, 55\penalty0 (2):\penalty0 1--36, 2022.

\bibitem[Jin et~al.(2023)Jin, Kim, Chen, Comeau, Yeganova, Wilbur, and Lu]{jin2023medcpt}
Qiao Jin, Won Kim, Qingyu Chen, Donald~C Comeau, Lana Yeganova, W~John Wilbur, and Zhiyong Lu.
\newblock Medcpt: Contrastive pre-trained transformers with large-scale pubmed search logs for zero-shot biomedical information retrieval.
\newblock \emph{Bioinformatics}, 39\penalty0 (11):\penalty0 btad651, 2023.

\bibitem[Jin et~al.(2024)Jin, Yang, Chen, and Lu]{jin2024genegpt}
Qiao Jin, Yifan Yang, Qingyu Chen, and Zhiyong Lu.
\newblock Genegpt: augmenting large language models with domain tools for improved access to biomedical information.
\newblock \emph{Bioinformatics}, 40\penalty0 (2):\penalty0 btae075, 2024.

\bibitem[Kadavath et~al.(2022)Kadavath, Conerly, Askell, Henighan, Drain, Perez, Schiefer, Hatfield-Dodds, DasSarma, Tran-Johnson, et~al.]{kadavath2022language}
Saurav Kadavath, Tom Conerly, Amanda Askell, Tom Henighan, Dawn Drain, Ethan Perez, Nicholas Schiefer, Zac Hatfield-Dodds, Nova DasSarma, Eli Tran-Johnson, et~al.
\newblock Language models (mostly) know what they know.
\newblock \emph{arXiv preprint arXiv:2207.05221}, 2022.

\bibitem[Karpukhin et~al.(2020)Karpukhin, Oguz, Min, Lewis, Wu, Edunov, Chen, and Yih]{karpukhin2020dense}
Vladimir Karpukhin, Barlas Oguz, Sewon Min, Patrick Lewis, Ledell Wu, Sergey Edunov, Danqi Chen, and Wen-tau Yih.
\newblock Dense passage retrieval for open-domain question answering.
\newblock In \emph{Proceedings of the 2020 conference on empirical methods in natural language processing (EMNLP)}, pages 6769--6781, 2020.

\bibitem[Krithara et~al.(2023)Krithara, Nentidis, Bougiatiotis, and Paliouras]{krithara2023bioasq}
Anastasia Krithara, Anastasios Nentidis, Konstantinos Bougiatiotis, and Georgios Paliouras.
\newblock Bioasq-qa: A manually curated corpus for biomedical question answering.
\newblock \emph{Scientific Data}, 10\penalty0 (1):\penalty0 170, 2023.

\bibitem[Lewis et~al.(2020)Lewis, Perez, Piktus, Petroni, Karpukhin, Goyal, K{\"u}ttler, Lewis, Yih, Rockt{\"a}schel, et~al.]{lewis2020retrieval}
Patrick Lewis, Ethan Perez, Aleksandra Piktus, Fabio Petroni, Vladimir Karpukhin, Naman Goyal, Heinrich K{\"u}ttler, Mike Lewis, Wen-tau Yih, Tim Rockt{\"a}schel, et~al.
\newblock Retrieval-augmented generation for knowledge-intensive nlp tasks.
\newblock \emph{Advances in neural information processing systems}, 33:\penalty0 9459--9474, 2020.

\bibitem[Li et~al.(2022)Li, Zhang, Ang, Ling, Sethi, Lee, Ginhoux, and Chen]{li2022disco}
Mengwei Li, Xiaomeng Zhang, Kok~Siong Ang, Jingjing Ling, Raman Sethi, Nicole Yee~Shin Lee, Florent Ginhoux, and Jinmiao Chen.
\newblock Disco: a database of deeply integrated human single-cell omics data.
\newblock \emph{Nucleic acids research}, 50\penalty0 (D1):\penalty0 D596--D602, 2022.

\bibitem[Li et~al.(2025)Li, Kilicoglu, Xu, and Zhang]{li2025biomedrag}
Mingchen Li, Halil Kilicoglu, Hua Xu, and Rui Zhang.
\newblock Biomedrag: A retrieval augmented large language model for biomedicine.
\newblock \emph{Journal of Biomedical Informatics}, 162:\penalty0 104769, 2025.

\bibitem[Lin(2004)]{lin2004rouge}
Chin-Yew Lin.
\newblock Rouge: A package for automatic evaluation of summaries.
\newblock In \emph{Text summarization branches out}, pages 74--81, 2004.

\bibitem[Luo et~al.(2025)Luo, Zhang, Yuan, Zhao, Yang, Gu, Wu, Chen, Qiao, Long, et~al.]{luo2025large}
Junyu Luo, Weizhi Zhang, Ye~Yuan, Yusheng Zhao, Junwei Yang, Yiyang Gu, Bohan Wu, Binqi Chen, Ziyue Qiao, Qingqing Long, et~al.
\newblock Large language model agent: A survey on methodology, applications and challenges.
\newblock \emph{arXiv preprint arXiv:2503.21460}, 2025.

\bibitem[Matsumoto et~al.(2024)Matsumoto, Moran, Choi, Hernandez, Venkatesan, Wang, and Moore]{matsumoto2024kragen}
Nicholas Matsumoto, Jay Moran, Hyunjun Choi, Miguel~E Hernandez, Mythreye Venkatesan, Paul Wang, and Jason~H Moore.
\newblock Kragen: a knowledge graph-enhanced rag framework for biomedical problem solving using large language models.
\newblock \emph{Bioinformatics}, 40\penalty0 (6):\penalty0 btae353, 2024.

\bibitem[Pal et~al.(2022)Pal, Umapathi, and Sankarasubbu]{pal2022medmcqa}
Ankit Pal, Logesh~Kumar Umapathi, and Malaikannan Sankarasubbu.
\newblock Medmcqa: A large-scale multi-subject multi-choice dataset for medical domain question answering.
\newblock In \emph{Conference on health, inference, and learning}, pages 248--260. PMLR, 2022.

\bibitem[Program et~al.(2025)Program, Abdulla, Aevermann, Assis, Badajoz, Bell, Bezzi, Cakir, Chaffer, Chambers, et~al.]{czi2025cz}
CZI Cell~Science Program, Shibla Abdulla, Brian Aevermann, Pedro Assis, Seve Badajoz, Sidney~M Bell, Emanuele Bezzi, Batuhan Cakir, Jim Chaffer, Signe Chambers, et~al.
\newblock Cz cellxgene discover: a single-cell data platform for scalable exploration, analysis and modeling of aggregated data.
\newblock \emph{Nucleic acids research}, 53\penalty0 (D1):\penalty0 D886--D900, 2025.

\bibitem[Qin et~al.(2025)Qin, Chen, Wang, Wu, Chen, Cheng, Zhao, Xiao, Dong, Long, et~al.]{qin2025scihorizon}
Chuan Qin, Xin Chen, Chengrui Wang, Pengmin Wu, Xi~Chen, Yihang Cheng, Jingyi Zhao, Meng Xiao, Xiangchao Dong, Qingqing Long, et~al.
\newblock Scihorizon: Benchmarking ai-for-science readiness from scientific data to large language models.
\newblock In \emph{Proceedings of the 31st ACM SIGKDD Conference on Knowledge Discovery and Data Mining V. 2}, pages 5754--5765, 2025.

\bibitem[Rajpurkar et~al.(2016)Rajpurkar, Zhang, Lopyrev, and Liang]{rajpurkar2016squad}
Pranav Rajpurkar, Jian Zhang, Konstantin Lopyrev, and Percy Liang.
\newblock Squad: 100,000+ questions for machine comprehension of text.
\newblock In \emph{Proceedings of the 2016 conference on empirical methods in natural language processing}, pages 2383--2392, 2016.

\bibitem[Shang et~al.(2025)Shang, Liao, Ji, and Hou]{shang2025benchmarking}
Xinyi Shang, Xu~Liao, Zhicheng Ji, and Wenpin Hou.
\newblock Benchmarking large language models for genomic knowledge with geneturing.
\newblock \emph{Briefings in Bioinformatics}, 26\penalty0 (5):\penalty0 bbaf492, 2025.

\bibitem[Shinn et~al.(2023)Shinn, Cassano, Gopinath, Narasimhan, and Yao]{shinn2023reflexion}
Noah Shinn, Federico Cassano, Ashwin Gopinath, Karthik Narasimhan, and Shunyu Yao.
\newblock Reflexion: Language agents with verbal reinforcement learning.
\newblock \emph{Advances in neural information processing systems}, 36:\penalty0 8634--8652, 2023.

\bibitem[Soman et~al.(2024)Soman, Rose, Morris, Akbas, Smith, Peetoom, Villouta-Reyes, Cerono, Shi, Rizk-Jackson, et~al.]{soman2024biomedical}
Karthik Soman, Peter~W Rose, John~H Morris, Rabia~E Akbas, Brett Smith, Braian Peetoom, Catalina Villouta-Reyes, Gabriel Cerono, Yongmei Shi, Angela Rizk-Jackson, et~al.
\newblock Biomedical knowledge graph-optimized prompt generation for large language models.
\newblock \emph{Bioinformatics}, 40\penalty0 (9):\penalty0 btae560, 2024.

\bibitem[Trivedi et~al.(2023)Trivedi, Balasubramanian, Khot, and Sabharwal]{trivedi2023ircot}
Harsh Trivedi, Niranjan Balasubramanian, Tushar Khot, and Ashish Sabharwal.
\newblock Interleaving retrieval with chain-of-thought reasoning for knowledge-intensive multi-step questions.
\newblock In \emph{Proceedings of the 61st annual meeting of the association for computational linguistics (volume 1: long papers)}, pages 10014--10037, 2023.

\bibitem[Wang et~al.(2024)Wang, Long, Xiao, Cai, Wu, Meng, Wang, and Zhou]{wang2024biorag}
Chengrui Wang, Qingqing Long, Meng Xiao, Xunxin Cai, Chengjun Wu, Zhen Meng, Xuezhi Wang, and Yuanchun Zhou.
\newblock Biorag: A rag-llm framework for biological question reasoning.
\newblock \emph{arXiv preprint arXiv:2408.01107}, 2024.

\bibitem[Wei et~al.(2024)Wei, Allot, Lai, Leaman, Tian, Luo, Jin, Wang, Chen, and Lu]{wei2024pubtator}
Chih-Hsuan Wei, Alexis Allot, Po-Ting Lai, Robert Leaman, Shubo Tian, Ling Luo, Qiao Jin, Zhizheng Wang, Qingyu Chen, and Zhiyong Lu.
\newblock Pubtator 3.0: an ai-powered literature resource for unlocking biomedical knowledge.
\newblock \emph{Nucleic Acids Research}, 52\penalty0 (W1):\penalty0 W540--W546, 2024.

\bibitem[Yao et~al.(2022)Yao, Zhao, Yu, Du, Shafran, Narasimhan, and Cao]{yao2022react}
Shunyu Yao, Jeffrey Zhao, Dian Yu, Nan Du, Izhak Shafran, Karthik Narasimhan, and Yuan Cao.
\newblock React: Synergizing reasoning and acting in language models.
\newblock \emph{arXiv preprint arXiv:2210.03629}, 2022.

\bibitem[Yu et~al.(2024{\natexlab{a}})Yu, Zhang, Pan, Cao, Ma, Li, Wang, and Yu]{yu2024chain}
Wenhao Yu, Hongming Zhang, Xiaoman Pan, Peixin Cao, Kaixin Ma, Jian Li, Hongwei Wang, and Dong Yu.
\newblock Chain-of-note: Enhancing robustness in retrieval-augmented language models.
\newblock In \emph{Proceedings of the 2024 conference on empirical methods in natural language processing}, pages 14672--14685, 2024{\natexlab{a}}.

\bibitem[Yu et~al.(2024{\natexlab{b}})Yu, Ping, Liu, Wang, You, Zhang, Shoeybi, and Catanzaro]{yu2024rankrag}
Yue Yu, Wei Ping, Zihan Liu, Boxin Wang, Jiaxuan You, Chao Zhang, Mohammad Shoeybi, and Bryan Catanzaro.
\newblock Rankrag: Unifying context ranking with retrieval-augmented generation in llms.
\newblock \emph{Advances in Neural Information Processing Systems}, 37:\penalty0 121156--121184, 2024{\natexlab{b}}.

\bibitem[Zhang et~al.(2024)Zhang, Jijo, Setty, Chung, Javid, Vidra, and Clifford]{zhang2024enhancing}
Liang Zhang, Katherine Jijo, Spurthi Setty, Eden Chung, Fatima Javid, Natan Vidra, and Tommy Clifford.
\newblock Enhancing large language model performance to answer questions and extract information more accurately.
\newblock \emph{arXiv preprint arXiv:2402.01722}, 2024.

\bibitem[Zhang et~al.(2026)Zhang, Gao, Tan, Zhou, Ding, Zhou, Zhang, and Wang]{zhang2026data}
Yunkun Zhang, Jin Gao, Zheling Tan, Lingfeng Zhou, Kexin Ding, Mu~Zhou, Shaoting Zhang, and Dequan Wang.
\newblock Data-centric foundation models in computational healthcare: A survey.
\newblock \emph{ACM Computing Surveys}, 58\penalty0 (11):\penalty0 1--35, 2026.

\bibitem[Zhou et~al.(2025)Zhou, Li, Chen, Chen, Han, and Gao]{zhou2025large}
Juexiao Zhou, Haoyang Li, Siyuan Chen, Zhangtianyi Chen, Zhongyi Han, and Xin Gao.
\newblock Large language models in biomedicine and healthcare.
\newblock \emph{npj Artificial Intelligence}, 1\penalty0 (1):\penalty0 44, 2025.

\end{thebibliography}

\end{document}